\documentclass[journal]{IEEEtran}
\usepackage{amsmath,epsfig}
\usepackage{amsmath,amssymb,amsfonts}
\usepackage{makecell}
\usepackage{url}
\usepackage{algorithmic}
\usepackage{graphicx}
\usepackage{textcomp}
\usepackage{cite}
\usepackage{times}
\usepackage{subfigure}
\usepackage{booktabs}
\usepackage{multirow}
\usepackage{caption}
\usepackage{color}
\newcommand{\blue}[1]{\textcolor{blue}{#1}}

\begin{document}
\title{SAR ATR Method with Limited Training Data via an Embedded Feature Augmenter and Dynamic Hierarchical-Feature Refiner}
%
%
%

\author{Chenwei Wang,~\IEEEmembership{Member,~IEEE,}
        Siyi Luo, 
        Yulin Huang,~\IEEEmembership{Senior Member,~IEEE,}\\
        Jifang Pei,~\IEEEmembership{Member,~IEEE,}
        Yin Zhang,~\IEEEmembership{Member,~IEEE,}
        Jianyu Yang,~\IEEEmembership{Member,~IEEE}
\thanks{This work was supported by \blue{the Sichuan Science and Technology Program under Grant 2023NSFSC1970}. (\emph{Corresponding author: Yulin Huang.})

       The authors are with the Department of Electrical Engineering, University of Electronic Science and Technology of China, Chengdu 611731, China (e-mail: yulinhuang@uestc.edu.cn; dbw181101@163.com).}}

\maketitle

\begin{abstract}
Without {sufficient} data, the quantity of information available for supervised training is constrained, {as obtaining sufficient synthetic aperture radar (SAR)} training data in practice {is frequently challenging}. Therefore, {current SAR automatic target recognition (ATR) algorithms perform poorly with limited training data availability, resulting in a critical need to increase SAR ATR performance.
In this study, a new method to improve SAR ATR when training data are limited is proposed.} 
First, an embedded feature augmenter is designed to enhance {the extracted} virtual features located far away from the class center. {Based on the relative distribution of the features, the algorithm pulls the corresponding virtual features with different strengths toward the corresponding class center.} The designed augmenter increases the amount of information available for supervised training and improves the separability of the extracted features.
Second, a dynamic hierarchical-feature refiner is proposed to capture {the} discriminative local features of the samples. Through dynamically generated kernels, the proposed refiner integrates {the} discriminative local features of different dimensions into {the} global features, {further enhancing} the inner-class compactness and inter-class separability of the extracted features.
The proposed method not only increases the amount of information available for supervised training but also {extracts the} discriminative features from the samples, resulting in superior ATR performance in problems with limited SAR training data. 
Experimental results on {the moving and stationary target acquisition and recognition (MSTAR)}, OpenSARShip, and FUSAR-Ship benchmark datasets demonstrate the {robustness} and outstanding ATR performance of the proposed method {in response to} limited SAR training data.
\end{abstract}

\begin{IEEEkeywords}
Synthetic aperture radar, automatic target recognition, embedded feature augmenter, dynamic hierarchical-feature refiner, limited training data
\end{IEEEkeywords}

%
\IEEEpeerreviewmaketitle

\section{Introduction}

\IEEEPARstart{S}{ynthetic} aperture radar (SAR) is an important microwave remote sensing system, renowned for its capability to acquire {high-resolution images regardless of weather or time of day \cite{intro1}.} In the field of SAR, automatic target recognition (ATR) is widely used {but remains challenging to implement} in SAR applications \cite{intro2}.
Over the past few decades, a wide range of recognition algorithms and systems have been proposed for SAR ATR. These methods can generally be grouped into three categories \cite{reff1,reff2,reff3,augment}, {namely,} template-based methods \cite{template2,wang2022recognition,template1}, model-based methods \cite{model2,model1}, and {deep-learning-based} methods \cite{ATR3,reff4,reff5,wang2023entropy,wang2023sar,reff6,ATR1,ATR2}.

Despite the promising performance demonstrated by current deep learning-based SAR ATR methods, a sufficient amount of information is still required for supervised training from a significant number of SAR images to avoid over-fitting and achieve excellent generalization performance across different imaging scenarios \cite{large1,wang2022global,large2,large3}.

In practice, {generating} a sufficient number of SAR images and labels for training {is resource-consuming} \cite{111,wang2022semi,intro_aug_dom2,wang2020deep,intro_method2}, which {can result} in {an} insufficient {amount of} information {available for properly} supervised training. {This scenario highlights a disconnection} between ATR methods and their practical {application} \cite{p3,p4,ATR1,p2}. 
In recent years, various methods have been proposed to address the problem of limited {amounts of} training images in SAR ATR \cite{add1, add2, add3}. {These methods share a common process—data augmentation, which is typically performed before model training. Some methods have proposed novel data augmentations, such as augmentations based on transformation consistency \cite{intro_aug_consit1,wang2022sar,open4}, domain knowledge \cite{intro_aug_dom1,wang2021multiview,intro_aug_dom2}, and generative adversarial networks \cite{intro_aug_GAN1,comparison1}. Traditional augmentation techniques have also been employed \cite{intro_method1,wang2019parking,intro_method2,intro_method3,comparison1,compared2}, such as flipping, cutting, and shifting \cite{open4,intro_aug2,wang2021deep,comparison3}.} 
Regardless of traditional or novel {augmentation}, two limitations {remain}: 1) additional data {preprocessing} or a separate generative model is required to generate {the} data {and} 2) It is hard to increase the total amount of information {for supervising} the training of {the} recognition model. 


However, for SAR ATR, {when the available training data is limited,} the information provided by {the} limited samples and corresponding labels alone is not sufficient to {provide} training of a recognition model with excellent performance and sufficient generalizability. According to several studies \cite{intro_key_instance1,wang2020multi,intro_key_instance2,li2023panoptic,liang2023efficient,intro_key_instance3}, information on the relative distributions of sample features and class centers can effectively improve the features {extracted by the recognition models}, increasing the total amount of information {used to supervise} the training of recognition {models}.

Therefore, to improve SAR ATR performance with limited training data, a novel method via an embedded feature augmenter and dynamic hierarchical-feature refiner is proposed. The outline is presented as follows and shown in Fig. \ref{framework}.
First, after obtaining the features of SAR images, the proposed embedded feature augmenter searches for features far away from the class center, i.e., {dissimilar inner-class} {and} similar inter-class feature pairs. {Then, the augmenter enhances the virtual features extracted from the features far away from the class center. Furthermore, an adaptive loss function is proposed to enhance inner-class compactness and inter-class separability by pulling the virtual features toward the corresponding class center with different strengths based on the relative distribution of the feature.} 

Second, a dynamic hierarchical-feature refiner is proposed. The dynamic hierarchical-feature refiner {first} enhances the class discriminability of local features {and} then dynamically generates convolutional kernels to integrate more discriminative local features of different dimensions into {the} global features. The kernels are dynamically generated based on the unique characteristics of the input SAR images. {Consequently}, the inner-class compactness and inter-class separability of the extracted features are further enhanced.

Through the above steps, the proposed method is able to {increase} the total information available for supervised training as well as integrate more discriminative features from the samples {into the training}, achieving accurate recognition with limited SAR training data. 
The main contributions of this {study} are summarized as follows:

(1) {The proposed embedded feature augmenter utilizes the relative feature distribution to increase the total information available for supervised training and improve the separability of the extracted features.}

(2) {The dynamic hierarchical-feature refiner places the discriminative features in a hierarchical structure and adapts to the analyzed SAR image by dynamically generating convolutional kernels, further enhancing the inner-class compactness and inter-class separability of the features.}

(3) The proposed method achieves state-of-the-art performance {in} recognition on three benchmark datasets {containing} limited training samples---the moving and stationary target acquisition and recognition (MSTAR), OpenSARship, and FUSAR-Ship {datasets}. Through ablation experiments, the {robustness} and effectiveness of the proposed method are validated as well.
 
The rest of this paper is organized as follows. Details of the method are presented in Section \uppercase\expandafter{\romannumeral2}.  The effectiveness of the proposed method is {demonstrated} by {experimental validation} in Section \uppercase\expandafter{\romannumeral3}, and {the conclusion is presented in Section \uppercase\expandafter{\romannumeral4}}.

\section{Proposed Method}
In this section, the proposed framework of the method is described in detail. {Then, the embedded feature augmenter and dynamic hierarchical-feature refiner are presented.} 


\subsection{Framework of Proposed Method}
Our method first extracts the low-dimensional feature maps from the limited training sample. Then, an embedded feature augmenter is implemented to achieve in-network augmentation of virtual features extracted from features far away from the class center. Finally, a dynamic hierarchical-feature refiner is used to place features in a hierarchical structure from which they are enhanced by adaptively generated kernels, ultimately producing improved recognition performance with limited training SAR data.

As shown in Fig. \ref{framework}, the framework of the proposed method consists of two main modules: {an} embedded feature augmenter and {a} dynamic hierarchical-feature refiner. First, we use a feature extractor to obtain the {features} of SAR images.
The feature extractor can be implemented by plain {convolutional neural networks} or transformers to extract low-dimensional feature maps from the limited training samples. 

The embedded feature augmenter is then implemented, consisting of two main steps: similarity pair search and virtual feature augmentation. In the first step, considering the two classes presented in Fig. \ref{framework}, the similarity pair search calculates the cosine similarity between each two samples to find features far away from the class center, i.e., the most dissimilar inner-class pairs and most similar inter-class pairs. Then, virtual features are generated from the features of these sample pairs, along with the corresponding class labels. Furthermore, an adaptive loss function $L_{ada}$ is applied to improve the separability of the extracted features. The adaptive loss pulls the virtual features closer to the corresponding class centers with different strengths. The corresponding strength of each sample is based on the relative distributions of the features, hence this loss is called the adaptive loss. In this way, features far away from the class center are subtly clustered or distanced with the adaptive loss $L_{ada}$. 
Therefore, the final loss is 
\begin{equation}
L_total= {{\lambda }_{1}}L_{reg} + {{\lambda }_{2}} L_{ada}   
\end{equation}
where ${\lambda_1}$ and ${\lambda_2}$ are the weighting coefficients. The model is updated with back-propagation by calculating the gradients based on $L_total$.

The dynamic hierarchical-feature refiner also consists of two main steps: local enhancement and global enhancement. {For sample sets that are limited, making accurate recognition difficult}, local enhancement calculates a spatial mask to {enhance} the class discriminability of local features, illustrated in step 1 of Fig. \ref{framework}. In step 2, the dynamic hierarchical-feature refiner integrates more discriminative local features of different dimensions into {the} global features {through} dynamically generated kernels.

Through the framework above, the embedded feature augmenter increases {the total amount of information available for} supervised training and improves the separability of {the features}. The dynamic hierarchical-feature refiner further enhances the inner-class
compactness and inter-class separability of {the features} in a hierarchical structure with {dynamically generated kernels}.
Ultimately, after optimization, our method can improve the recognition performance of SAR ATR with the {use of} limited training SAR samples.

\subsection{Embedded Feature Augmenter}
The embedded feature augmenter aims to {utilize} the information from {the relative distribution} of {the features} for supervised training {to improve} the separability of the extracted features. As shown in Fig. \ref{SMixup}, the embedded feature augmenter mainly consists of two stages: {1) similarity pair search and 2) virtual feature augmentation.} The adaptive loss is shown in the right orange block.

Given input SAR images $\left\{ \mathbf{x}_{1}^{1},\mathbf{x}_{2}^{1},\ldots ,\mathbf{x}_{B}^{C} \right\}$, ${{\mathbf{x}}_j^i}$ is the $j\rm{th}$ SAR image {in class $i$.} The SAR images are first {input} into the feature extractor to obtain the feature maps. 
Then, the first stage, similarity pair search, is performed to locate features far away from the class center, i.e., the most dissimilar inner-class pairs and most similar inter-class pairs. The $j\rm{th}$ SAR image sample {of the $i\rm{th}$ class}, ${{\mathbf{x}}_j^i}$ is set as an example. The similarities between ${{\mathbf{x}}_j^i}$ and {the} other samples are calculated by

\begin{figure}[thb]
\centering
\includegraphics[width=0.95\linewidth]{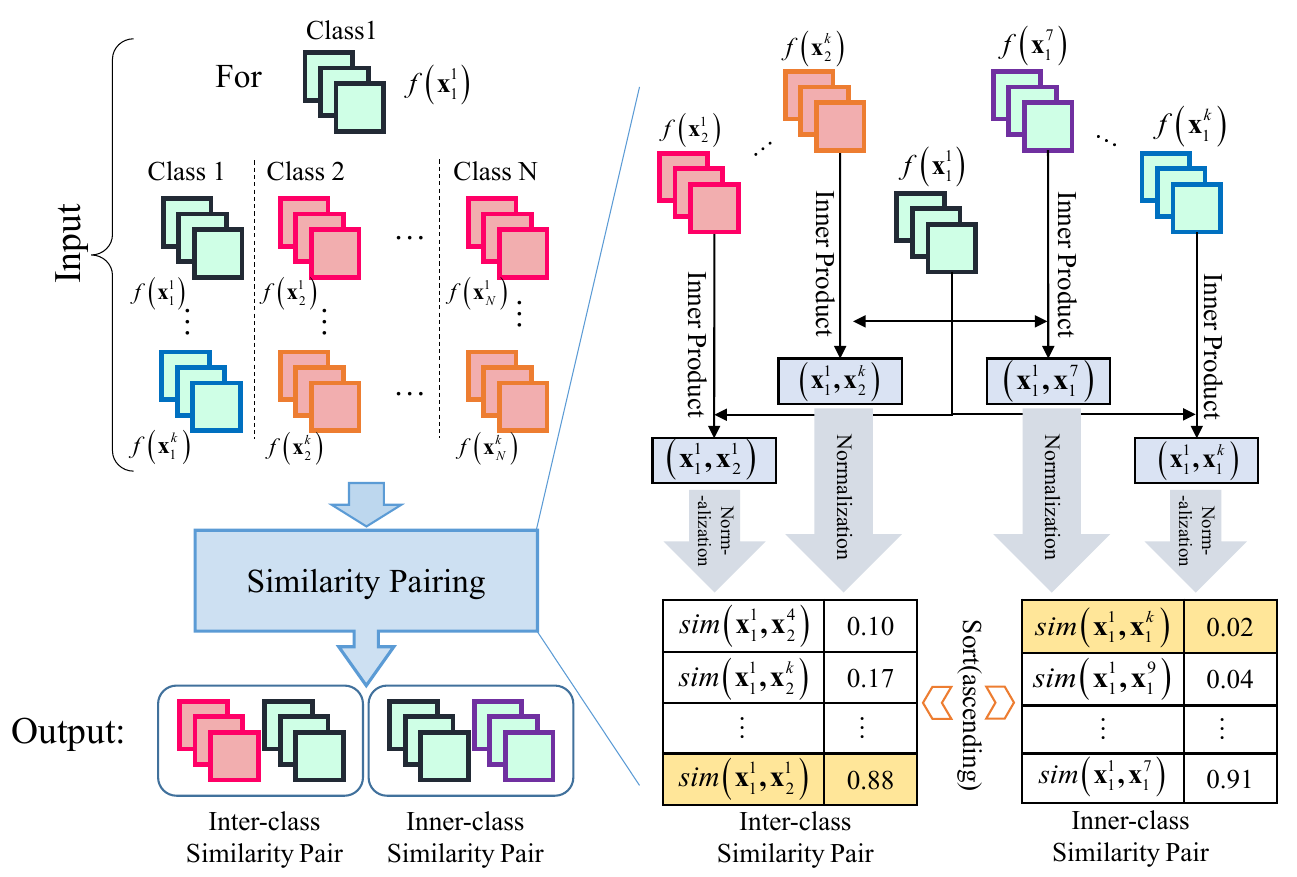} 
\caption{Details of similarity calculation in step 1 of {the} embedded feature augmenter. The inter-class similarity pairing {finds} the most similar inter-class pair, {and} the inner-class similarity {pairing finds} the most dissimilar inner-class pair.}
\label{similarity}
\end{figure}

\begin{equation}
\label{eq1}
sim\left ( {{\mathbf{x}}_j^i},\mathbf{x}_k^i \right ) =\frac{f\left ({{\mathbf{x}}_j^i}\right )}{\left \| f\left ({{\mathbf{x}}_j^i}\right ) \right \| _{2}}\cdot \frac{f\left (\mathbf{x}_k^i\right )}{\left \| f\left (\mathbf{x}_k^i\right ) \right \|_{2} }
\end{equation}
where $ {\left\|  \cdot  \right\|_2} $ {refers to} the $\rm{L}_2$-norm and $f(\mathbf{x}_k^i)$ {refers to} the feature maps of $\mathbf{x}_k^i$ after {passing} through the feature extractor. {Accordingly,} for ${{\mathbf{x}}_j^i}$, the inner-class dissimilar sample and inter-class similar sample can be found. {An inner-class dissimilar pair and an inter-class similar pair are identified and shown as red lines in subpart C of stage 1 in Fig. \ref{SMixup}.}

The second stage, virtual feature {augmentation, is} based on the constructed pairs. {Virtual feature augmentation} uses a random value to control the weights of the feature maps in the pair. The ground truth of the new feature maps is also {a mix-up} from the ground truths of the pair with the same random value. The process is shown in subpart A of stage 2 in Fig. \ref{SMixup} and can be formulated as

\begin{equation}
g\left (\mathbf{x}_k^i,\mathbf{x}_l^j \right )=\alpha f\left (\mathbf{x}_k^i \right )+\left (1-\alpha \right ) f\left (\mathbf{x}_l^j \right )
\end{equation}
where $g\left (\mathbf{x}_k^i,\mathbf{x}_l^j \right )$ {refers to} the virtual feature between the {pair} $\mathbf{x}_k^i$ and $\mathbf{x}_l^j$ {and $\alpha \in \left [ 0,1 \right ] $ is a random value that fits the distribution.} Finally, the virtual feature {is} concatenated with the {pair, and the feature maps go through the classifier, calculating the recognition loss as}

\begin{equation}
L_{reg}=-\sum_{i=1}^{C}\sum_{j=1}^{B}y_i\mathit{log}\left ( p\left ( y_i|\mathbf{x}_j^i  \right )  \right )   
\end{equation}
where the predicted possibility ${{p}}\left( {{y}_{i}}|{{\mathbf{x}}_{j}^{i}} \right)$ is output by the classifier with a SoftMax layer {and} $B$ is the SAR image number of each class.

As shown in the right orange block of Fig. \ref{SMixup}, for ${{\mathbf{x}}_j^i}$, {suppose the similar inter-class feature, dissimilar inner-class feature, well-recognized inter-class feature, and well-recognized inner-class feature are ${{\mathbf{x}}_{inter}^{hard}}$, ${{\mathbf{x}}_{inner}^{hard}}$, ${{\mathbf{x}}_{inter}^{easy}}$, and ${{\mathbf{x}}_{inner}^{easy}}$, respectively.}
We first find the corresponding well-recognized samples of ${{\mathbf{x}}_j^i}$ based on Eq. (\ref{eq1}). Then, the benchmark distance of well-recognized {samples} can be calculated as 
\begin{align}
d_{bm}=sim\left ( {\bf{x}}_j^i,{\bf{x}}_{inter}^{easy} \right ) -sim\left ( {\bf{x}}_j^i,{\bf{x}}_{inner}^{easy} \right )
\end{align}

Then, the dynamic intensities can be calculated as 
\begin{equation}
\begin{aligned}
{{\psi }_{dyn}}\left( \mathbf{x}_{j}^{i} \right)= & \left( 1- \right.\left( sim_{inner}^{hard}\left( \mathbf{x}_{j}^{i} \right) \right. \\ 
 & {{\left. \left. -sim_{\operatorname{int}er}^{hard}\left( \mathbf{x}_{j}^{i} \right)+2 \right)/4 \right)}^{\rho }}*{{d}_{bm}} \\ 
\end{aligned}
\end{equation}
where $\psi _{dyn}\left ( {\bf{x}}_1^1 \right )$ is the dynamic intensity for ${\bf{x}}_1^1 $ {and} $\rho \ge 0$ is the adaptive parameter {used} to adjust the value of $d_{bm}$. The adaptive loss {function} is presented as 
\begin{equation}
\begin{aligned}
&& {{L}_{ada}}=\sum{\max } \left( sim\left( \mathbf{x}_{j}^{i},\mathbf{x}_{inter}^{hard} \right)+{{\psi }_{dyn}}\left( \mathbf{x}_{j}^{i} \right) \right. \\ 
 && \left. -sim\left( \mathbf{x}_{j}^{i},\mathbf{x}_{inner}^{hard} \right),0 \right) \\
\end{aligned}
\end{equation}
where ${L}_{ada}$ is the adaptive loss.
In this way, for each feature pair, the adaptive loss can provide a dynamic intensity based on the difference between the distance of one hard-to-recognize {pair} and the benchmark distance $d_{bm}$. 

As shown in subpart B of stage 2 in Fig. \ref{SMixup}, after optimization, {the embedded feature augmenter can utilize the information from the relative distribution of the features to supervise the training and improve the separability of the features. Furthermore, from a theoretical perspective, these constraints on the features serve as regularization on the hypothesis space,} leading to a smaller hypothesis space that is effective and crucial for SAR ATR under limited training samples \cite{Generalizing}.

\subsection{Dynamic Hierarchical-Feature Refiner}
\begin{figure}[thb]
\centering
\includegraphics[width=0.98\linewidth]{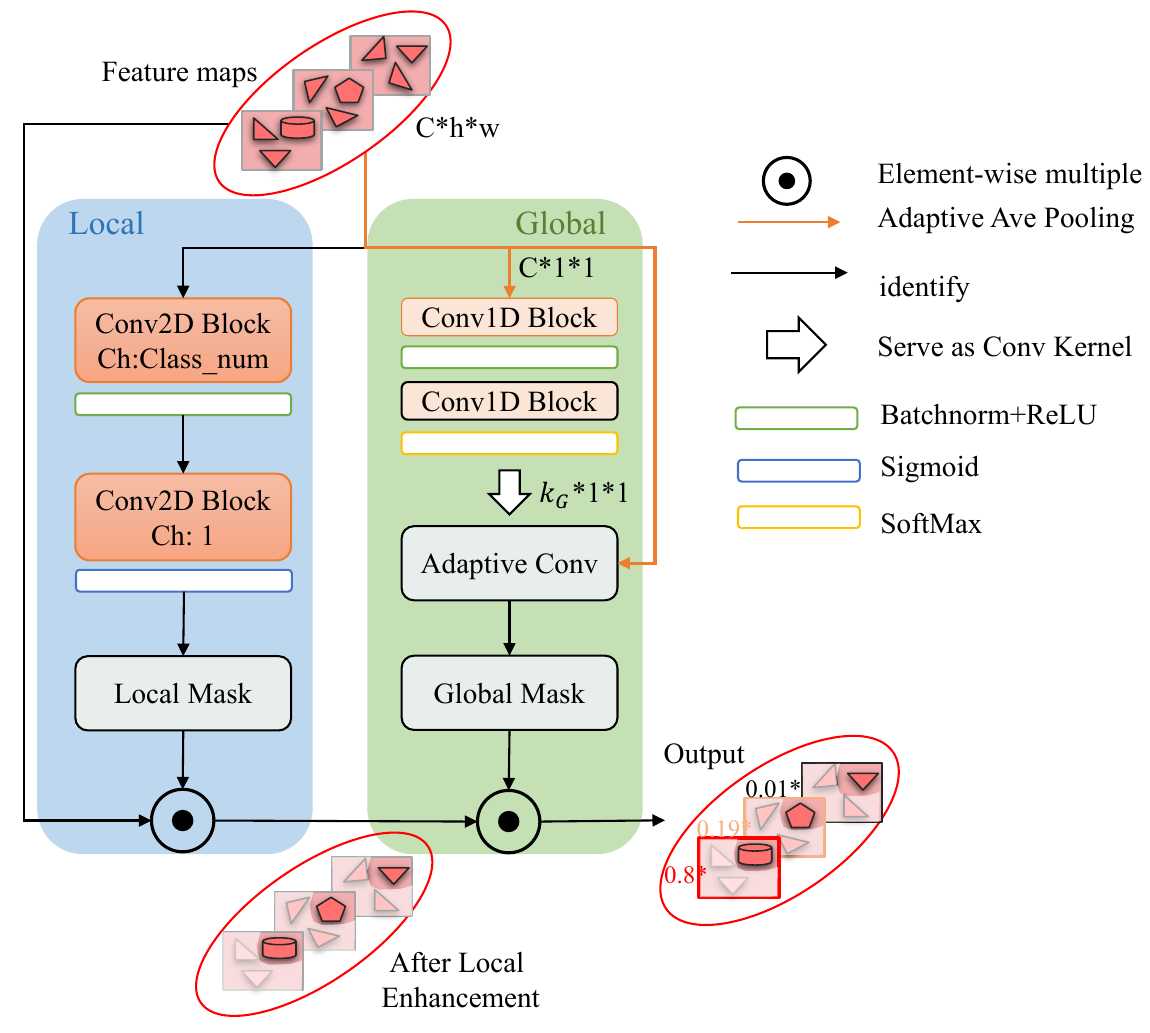} 
\caption{Dynamic hierarchical-feature refiner. Dynamic hierarchical-feature refiner consists of two stages: local enhancement (blue box) and global enhancement (green box). {The output feature maps show the enhancement of the spatial features by the local mask. The different colors of the borders refer to the different weighting coefficients of the feature maps; brighter colors correspond to larger coefficients.}}
\label{Glocal}
\end{figure}

The dynamic hierarchical-feature refiner aims to further {enhance} the inner-class
compactness and inter-class separability of {features} by {classifying the} more discriminative features in a hierarchical structure {to be applied} to the specific SAR image by dynamically generating kernels. The dynamic hierarchical-feature refiner consists of two stages: local enhancement and global enhancement as shown in Fig. \ref{Glocal}.

Given the feature maps of one sample, $f\left (\mathbf{x}_k^i\right )$, the first stage, local enhancement, {uses} a local mask to enhance the feature maps from the spatial aspect. One conv2D block with {batch normalization (BN) \cite{batchnorm} and the ReLu activation function \cite{ReLu}} first refines $f\left (\mathbf{x}_k^i\right )$ with fewer channels. {Then, another conv2D block with the Sigmoid activation function} produces the local mask. The local enhanced feature maps are obtained by element-wise {multiplication of the} local mask with $f\left (\mathbf{x}_k^i\right )$. The process of local enhancement can be {presented} as

\begin{equation}
L_k^i=f\left (\mathbf{x}_k^i\right )\odot f_S^{2D} \left (f_{bnr}^{2D}\left  (f\left (\mathbf{x}_k^i\right ),C\right )\right )
\end{equation}
where $\odot $ refers to the Hadamard product, $f_{bnr}^{2D} \left (\cdot \right )$ denotes conv2D with BN and ReLu, $C$ is the channel number of $f_{bnr}^{2D} \left (\cdot \right )$, and $f_S^{2D}\left (\cdot \right )$ refers to the conv2D block with Sigmoid.

In the second stage, global enhancement, a global mask is generated starting with $f\left (\mathbf{x}_k^i\right )$ to enhance the feature maps from the channel aspect. The global vector of $f\left (\mathbf{x}_k^i\right )$ {is obtained by adaptive average pooling}. The adaptive convolutional kernels $k_{gen}$ are calculated from the global vector by {passing through one conv1D block with BN and ReLu and one conv1D block with the SoftMax activation function}.  Then, the feature maps $f\left (\mathbf{x}_k^i\right )$ {are} convoluted with the generated kernels to obtain the final global mask. The process of global enhancement can be described as 

\begin{equation}
{GL}_k^i=L_k^i\odot \left (f\left (\mathbf{x}_k^i\right )*k_{gen}\right )
\end{equation}
where ${GL}_k^i$ denotes the final output of the dynamic hierarchical-feature refiner {and $*$ denotes the convolution function}. The generation process of the adaptive convolutional kernels can be presented as

\begin{equation}
k_{gen}=sf\left (f^{1D}\left ( f_{bnr}^{1D}\left  (AAP\left (f\left (\mathbf{x}_k^i\right ) \right )\right )\right )\right )
\end{equation}
where $sf\left (\cdot \right )$ {refers to} the SoftMax function, $f^{1D} \left (\cdot \right )$ {denotes} the conv1D block, $f_{bnr}^{1D}\left (\cdot \right )$ {is} the conv1D block with BN and ReLu, and $AAP\left (\cdot \right )$ {denotes} the adaptive average pooling.

Through {local and global enhancement}, the dynamic hierarchical-feature refiner integrates more discriminative features. {Meanwhile, it can be regarded as an attention mechanism, providing the model with an intrinsic data structure that can improve the search strategy of the model for the optimal hypothesis \cite{Generalizing}.}

The scattering characteristic of SAR images has complex dynamics {owing to changes in the imaging platform, azimuth, and other features} \cite{intro_sar_complex1, intro_sar_complex2}. {Conventional attention methods, such as SE \cite{SE} and CBAM \cite{CBAM}, use kernels that are fixed after optimization, may lack sufficient representation power for complex dynamics in SAR images.} Different from these attention methods, {our hierarchical-feature refiner dynamically calculates kernels based on different SAR images to handle the changing scattering characteristics of SAR images}.

The proposed method invokes an embedded feature augmenter to increase the total amount of information for supervised training and separability of features and designs a dynamic hierarchical-feature refiner to integrate more discriminative features by extracting the local and global specific intrinsic structure in the data. Next, we present experiments conducted with limited data to validate the effectiveness and practicability of the proposed method.

\section{Experiments and Results}

In this section, we first introduce the MSTAR, OpenSARShip, and FUSAR-Ship datasets--—three benchmark datasets for SAR ATR. The ablation experiments are evaluated with three different sets of training samples.
Then, we run experiments on {each of} the three datasets, with the training samples for each class {representing} a wide range, to validate the effectiveness of our method under different degrees of limited training data.
Finally, {comparisons with other methods under different datasets are presented.}

\subsection{Dataset and Preprocessing}

Three benchmark datasets are used to validate the effectiveness and robustness of our methods. They are MSTAR, OpenSARShip, and FUSAR-Ship datasets.

\begin{figure}
\centering
\includegraphics[width=0.48\textwidth]{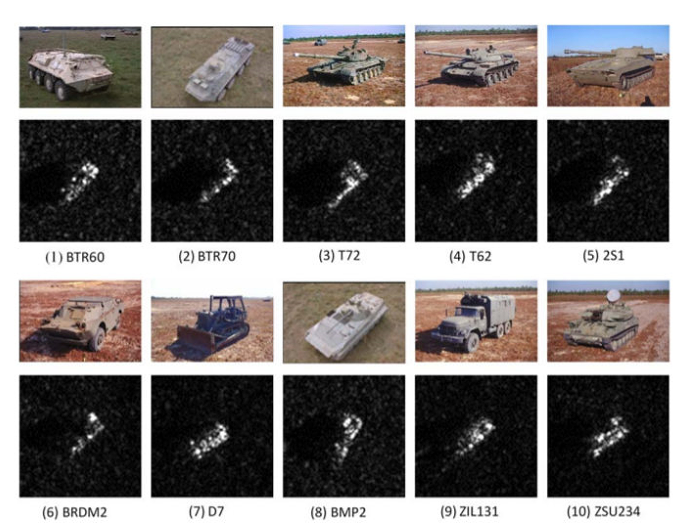}
\caption{SAR images and corresponding optical images of targets.}
\label{sampleMSTAR}
\end{figure}

As a benchmark dataset for SAR ATR performance evaluation, MSTAR is collected by the Sandia National Laboratory STARLOS sensor platform and published by the Defense Advanced Research Projects Agency and Air Force Research Laboratory. This dataset {contains} a large number of SAR images with various types of military vehicles and clutter images. {Represented in these images are 10 different classes of ground targets, such as tanks, rocket launchers, armored personnel carriers, air defense units, and bulldozers, acquired as 1-foot resolution X-band SAR images with full aspect coverage (ranging from $0^{\circ}$ to $360^{\circ}$).} In addition, they are captured under different depression angles and serial numbers. The SAR and corresponding optical images of {these 10 targets at similar azimuth angles are shown in Fig. \ref{sampleMSTAR}.}


\begin{figure}[thb]
\centering
\includegraphics[width=1\linewidth]{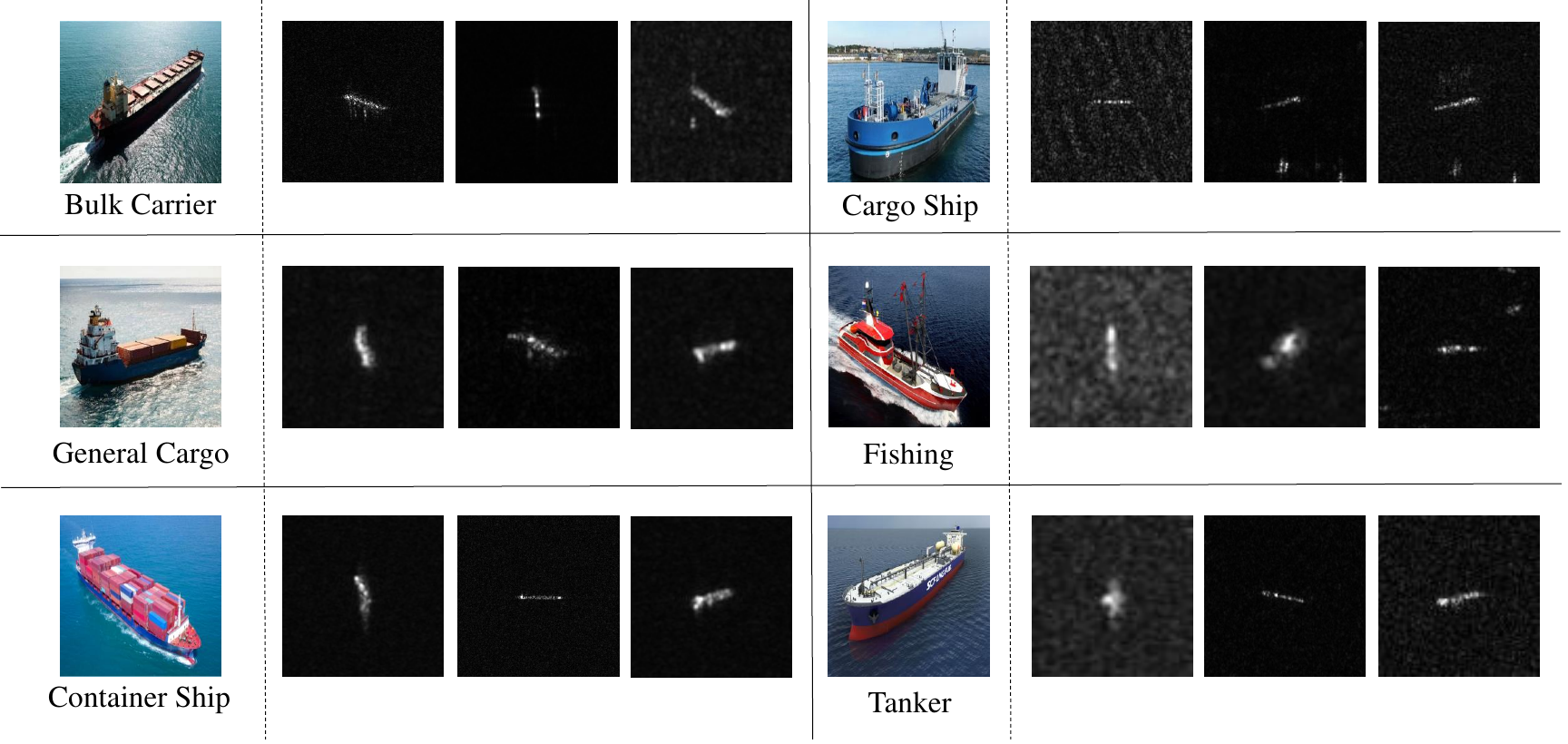} 
\caption{SAR ship images and corresponding optical ship images of six ship classes in {the} OpenSARShip dataset.}
\label{sampleOPEN}
\end{figure}

\begin{figure}[thb]
\centering
\includegraphics[width=1\linewidth]{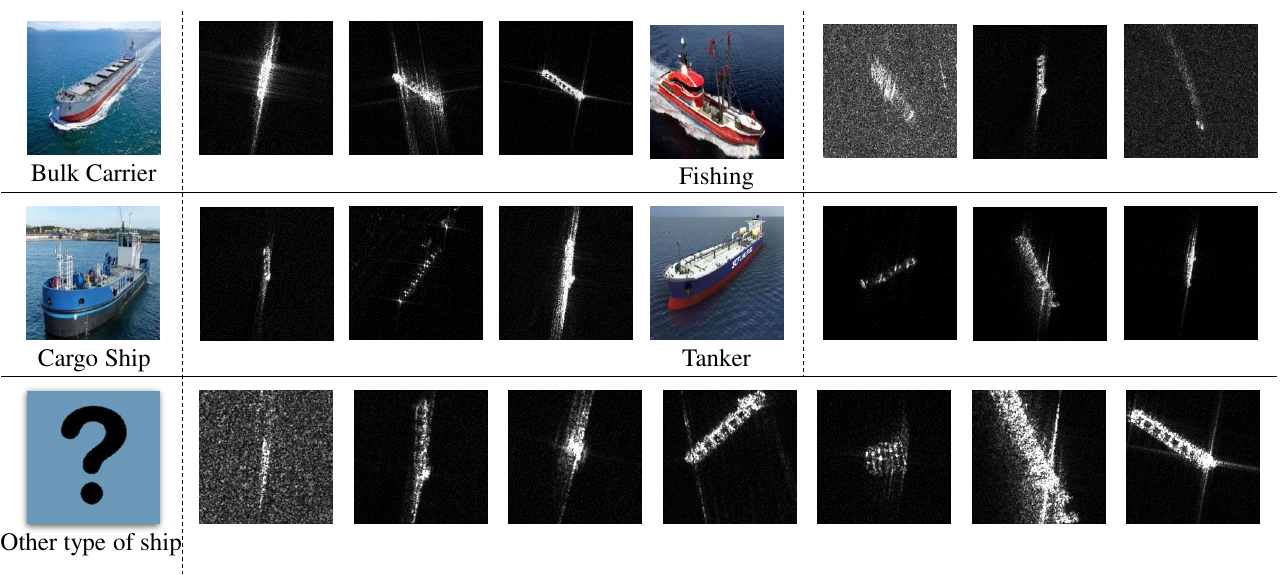} 
\caption{SAR ship images and corresponding optical ship images of five ship classes in {the} FUSAR-Ship dataset. {The last ship class, other types of ships, is a class that includes most of the other ship classes apart from the common ship classes.} This ship class has more overlap with other ship classes and can validate the robustness and effectiveness of our method more comprehensively.}
\label{sampleFUSAR}
\end{figure}

To develop sophisticated ship detection and classification algorithms, the OpenSARShip dataset is collected from 41 Sentinel-1 images under different environmental conditions. {The dataset covers 17 types of ships, totaling 11,346 SAR ships combined with automatic identification system (AIS) information. In our experiment, ground range detection was selected in Sentinel-1 IW mode with a resolution of $2.0 m \times 1.5 m$ and pixel size of $10 m \times 10 m$ in the azimuth and range directions, respectively.} The length of the ships ranges from 92 to 399 m, while the width ranges from 6 to 65 m. {Notably, the annotation of this dataset is highly reliable, as it is based on AIS information \cite{ais}.} We decided to use VV and VH data in the training, validation, and testing process. Fig.\ref{sampleOPEN} shows the SAR images of these types of targets from the OpenSARShip dataset.

Another open benchmark dataset for ship and marine target detection and recognition, FUSAR-Ship \cite{FUSAR}, is compiled by the Key Lab of Information Science of Electromagnetic Waves of Fudan University {from the Gaofen-3 (GF-3) satellite. The GF-3 satellite} is the first civilian C-band fully polarized satellite-based SAR in China, mainly used for marine remote sensing and marine monitoring. {As open SAR-AIS matching dataset,} FUSAR-Ship is constructed for more than 100 GF-3 scenes, including over 5000 ship image slices with AIS information. 

The FUSAR-Ship dataset is a new dataset {with} four different imaging parameters {from the} OpenSARShip dataset, including incident angle, bandwidth, and resolution. The ranges of the four imaging parameters in {the} FUSAR-Ship dataset are larger than {those of} OpenSARShip. Although {the} FUSAR-Ship dataset may achieve a higher resolution of $0.5$m, these larger ranges make {the} FUSAR-Ship dataset more complex than OpenSARShip for recognition. 

The OpenSARShip and FUSAR-Ship datasets can be described by three main features, which include incident angle, bandwidth, and resolution. The configurations of the training and the network are presented as follows. The input SAR images are of size $224\times 224$. 
The tiny version of the swin transformer \cite{swin} is employed as the feature extractor. 
The size of input SAR images is $224 \times 224$ by applying bi-linear interpolation to the original data.
The value of $\lambda_1$ and $\lambda_2$ is set as $1$ and $0.8$, which is selected according to \cite{arlot2010survey}.
The SGD optimizer is employed in training, the value of weight decay is set as $5\times10^{-5}$, the value of the momentum is set as $0.9$.
The random value $\alpha $ obeys the beta distribution, in which $\alpha $ and $\beta $ are set to 0.1. 
The epoch number of the training process is set to 160. 
The batch size is set to 32, and the learning rate is initialized as 0.01 and is reduced by cosine annealing with a periodicity of 320. There are also 15 epochs used for training warmup.
The proposed method is tested and evaluated on a GPU cluster with Intel(R) Xeon(R) CPU E5-2698 v4 @ 2.20GHz, eight Tesla V100 with eight 32GB memories. The proposed method is implemented using the open-source PyTorch framework with only one Tesla V100.


\renewcommand{\arraystretch}{1.2}
\begin{table}[]
\centering
\caption{Dataset: Raw Images in MSTAR Dataset Under SOC}
\setlength\tabcolsep{7pt}
\label{ttnumMSTAR}
\begin{tabular}{lccccc}
\hline\hline
Target Type & BMP2 & BRDM2 & BTR60 & BTR70 & D7 \\ \hline
 Training(17\textdegree) & 233 & 298 & 256 & 233 & 299 \\
Testing(15\textdegree) & 195 & 274 & 195 & 196 & 274 \\ \hline\hline
Target Type & 2S1 & T62 & T72 & ZIL131 & ZSU235 \\ \hline
Training(17\textdegree) & 299 & 299 & 232 & 299 & 299 \\
Testing(15\textdegree) & 274 & 273 & 196 & 274 & 274 \\ \hline\hline
\end{tabular}
\end{table}

\begin{figure}[]
\begin{center}
\subfigure[V0]{\label{a.0}\includegraphics[width=0.23\textwidth]{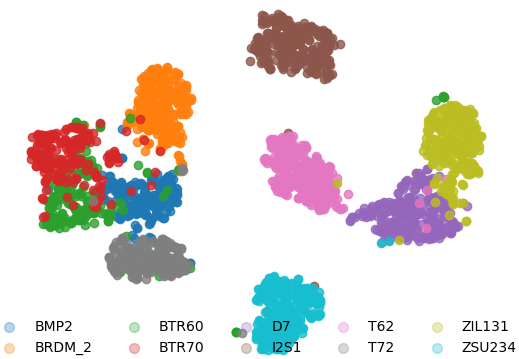}}
\subfigure[V1]{\label{a.1}\includegraphics[width=0.23\textwidth]{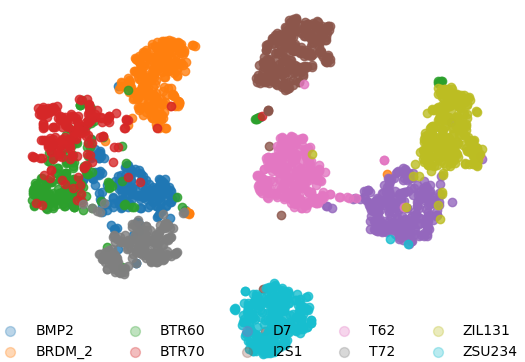}} \\
\subfigure[V2]{\label{a.2}\includegraphics[width=0.23\textwidth]{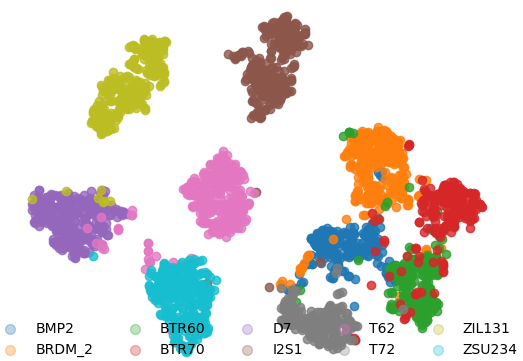}}
\subfigure[V3]{\label{a.3}\includegraphics[width=0.23\textwidth]{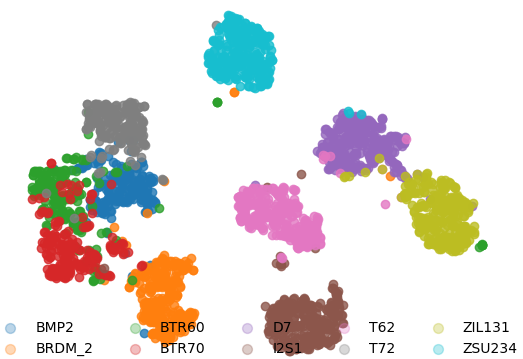}} \\
\subfigure[V4]{\label{a.4}\includegraphics[width=0.23\textwidth]{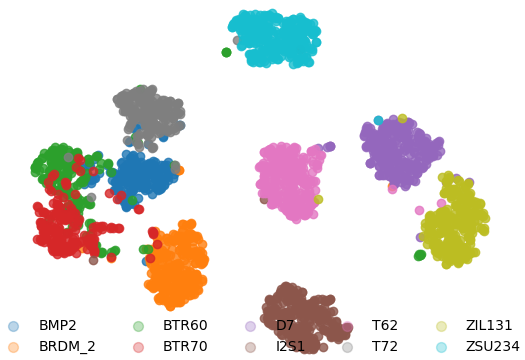}} 
\subfigure[Ours]{\label{1.4}\includegraphics[width=0.23\textwidth]{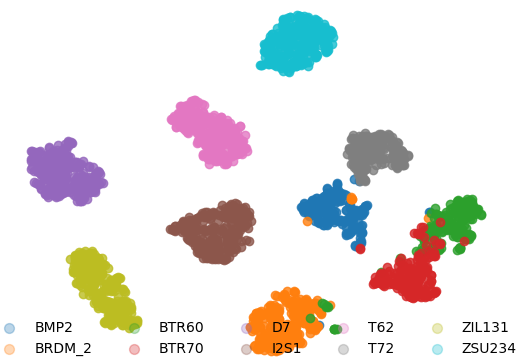}}\\
\end{center}
\caption{Feature visualization of different configurations with 20 each class under MSTAR. The configurations are the same as Table \ref{ablation_tab}}
\label{ablation_fig}
\end{figure}

\renewcommand{\arraystretch}{1.6}
\begin{table*}[h]
\centering
\footnotesize
\caption{Ablation experiments: Recognition performance (\%) of different ablation configurations under different numbers of training samples. EFA , Ada Loss, and DHFR refer to the embedded feature augmenter, adaptive loss, and dynamic hierarchical-feature refiner, respectively.}
\label{ablation_tab}
\begin{tabular}{c|c|ccc|ccccccccccc}
\hline \hline
\makebox[21pt][c]{\multirow{2}{*}{\begin{tabular}[c]{@{}c@{}}Training\\ Number\end{tabular}}} &
  \makebox[21pt][c]{\multirow{2}{*}{Method}} &
  \makebox[21pt][c]{\multirow{2}{*}{EFA}} &
  \makebox[21pt][c]{\multirow{2}{*}{Ada Loss}} &
  \makebox[21pt][c]{\multirow{2}{*}{DHFR}} &
  \makebox[19.8pt][c]{\multirow{2}{*}{BMP2}} &
  \makebox[19.8pt][c]{\multirow{2}{*}{BRDM2}} &
  \makebox[19.8pt][c]{\multirow{2}{*}{BTR60}} &
  \makebox[19.8pt][c]{\multirow{2}{*}{BTR70}} &
  \makebox[19.8pt][c]{\multirow{2}{*}{D7}} &
  \makebox[19.8pt][c]{\multirow{2}{*}{2S1}} &
  \makebox[19.8pt][c]{\multirow{2}{*}{T62} }&
  \makebox[19.8pt][c]{\multirow{2}{*}{T72}} &
  \makebox[19.8pt][c]{\multirow{2}{*}{ZIL131}} &
  \makebox[19.8pt][c]{\multirow{2}{*}{ZSU234}} &
  \makebox[19.8pt][c]{\multirow{2}{*}{Average}} \\
                     &      &   &   &   &       &       &       &       &       &       &       &       &       &       &       \\ \hline 
\multirow{6}{*}{100} & V0   & × & × & × & 78.97 & 87.59 & 77.95 & 84.18 & 81.39 & 85.77 & 97.07 & 95.92 & 95.26 & 99.64 & 88.91 \\
    & V1 & \checkmark & × & × & 87.18 & 94.42 & 87.18 & 98.47 & 91.97 & 85.40 & 91.21 & 98.47 & 98.91 & 99.64 & 93.35 \\
    & V2   & \checkmark & \checkmark & × & 85.64 & 93.43 & 86.67 & 90.31 & 89.05 & 90.51 & 99.27 & 99.49 & 99.64 & 99.64 & 93.73 \\
                     & V3   & \checkmark & × & \checkmark & 89.74 & 98.91 & 93.33 & 92.86 & 99.64 & 94.89 & 99.27 & 97.96 & 98.18 & 99.64 & 96.82 \\
                     & V4   & × & × & \checkmark & 90.26 & 92.70 & 79.49 & 88.78 & 82.48 & 91.61 & 89.74 & 98.47 & 98.18 & 99.64 & 91.38 \\
                     & Ours & \checkmark & \checkmark & \checkmark & 98.97 & 98.54 & 98.97 & 99.49 & 99.64 & 99.64 & 99.63 & 99.49 & 99.64 & 99.64 & 99.38 \\ \hline 
\multirow{6}{*}{60} & V0   & × & × & × & 78.46 & 95.62 & 83.59 & 92.86 & 78.83 & 62.77 & 81.32 & 96.94 & 97.45 & 99.64 & 86.60 \\
    & V1 & \checkmark & × & × & 89.74 & 93.07 & 85.64 & 93.88 & 98.18 & 90.51 & 98.17 & 96.43 & 97.81 & 99.64 & 94.68 \\
    & V2   & \checkmark & \checkmark & × & 90.77 & 99.27 & 90.26 & 89.80 & 99.64 & 94.16 & 91.58 & 96.43 & 97.08 & 99.64 & 95.26 \\
                     & V3   & \checkmark & × & \checkmark & 90.77 & 98.54 & 88.21 & 91.33 & 97.81 & 90.15 & 97.44 & 94.90 & 97.08 & 99.64 & 95.01 \\
                     & V4   & × & × & \checkmark & 82.05 & 93.80 & 85.64 & 95.41 & 85.04 & 79.56 & 94.87 & 97.45 & 98.54 & 99.64 & 91.34 \\
                     & Ours & \checkmark & \checkmark & \checkmark & 98.97 & 98.18 & 97.44 & 99.49 & 99.64 & 98.54 & 99.63 & 99.49 & 99.64 & 99.64 & 99.09 \\ \hline 
\multirow{6}{*}{20}  & V0   & × & × & × & 72.82 & 86.13 & 67.18 & 79.59 & 54.74 & 77.74 & 73.99 & 85.20 & 99.27 & 99.64 & 80.08 \\
    & V1 & \checkmark & × & × & 76.92 & 94.53 & 77.95 & 85.20 & 77.01 & 82.12 & 86.45 & 94.39 & 99.64 & 99.64 & 87.88 \\
    & V2   & \checkmark & \checkmark & × & 79.49 & 97.08 & 80.00 & 85.20 & 78.10 & 88.32 & 79.49 & 92.86 & 99.64 & 99.64 & 88.45 \\
                     & V3   & \checkmark & × & \checkmark & 84.62 & 96.35 & 78.97 & 84.69 & 79.56 & 86.50 & 84.98 & 91.84 & 99.27 & 99.64 & 89.11 \\
                     & V4   & × & × & \checkmark & 87.18 & 92.70 & 70.26 & 84.69 & 64.60 & 78.10 & 86.45 & 91.84 & 99.27 & 99.64 & 85.73 \\
                     & Ours & \checkmark & \checkmark & \checkmark & 90.77 & 98.54 & 94.36 & 94.90 & 94.16 & 94.89 & 99.27 & 99.49 & 99.64 & 99.64 & 96.78 \\ \hline \hline 
\end{tabular} 
\end{table*}

\renewcommand{\arraystretch}{1.6}
\begin{table}[h]
\centering
\footnotesize
\caption{Ablation experiments: Recognition performance (\%) of different ablation configurations under different numbers of training samples in OpenSARShip. EFAA  and DHFR stand for embedded feature augmenter with adaptive loss and dynamic hierarchical-feature refiner, respectively.}
\label{ablation_tab2}
\setlength{\tabcolsep}{3.5pt}
\begin{tabular}{c|c|cc|cccc}
\hline \hline 
\multirow{2}{*}{\begin{tabular}[c]{@{}c@{}}Training\\      Number\end{tabular}} &
  \multirow{2}{*}{Method} &
  \multirow{2}{*}{EFAA} &
  \multirow{2}{*}{DHFR} &
  \multirow{2}{*}{\begin{tabular}[c]{@{}c@{}}Bulk\\Carrier\end{tabular}} &
  \multirow{2}{*}{\begin{tabular}[c]{@{}c@{}}Container\\Ship\end{tabular}} &
  \multirow{2}{*}{Tanker} &
  \multirow{2}{*}{Average} \\ 
                    &      &   &   &       &       &       &       \\ \hline 
\multirow{4}{*}{10} & V0   & × & × & 57.26 & 54.13 & 47.46 & 53.60 \\
                    & V2   & × & \checkmark  & 51.37 & 71.17 & 57.27 & 61.59 \\
                    & V4   & \checkmark  & × & 40.63 & 78.90 & 50.88 & 60.06 \\
                    & Ours & \checkmark  & \checkmark  & 48.84  & 77.19  & 71.47  & 67.74  \\ \hline
\multirow{4}{*}{20} & V0   & × & × & 68.27 & 62.17 & 45.49 & 57.93 \\
                    & V2   & × & \checkmark  & 71.79 & 67.37 & 56.61 & 65.67 \\
                    & V4   & \checkmark  & × & 76.63 & 49.57 & 70.06 & 61.83 \\
                    & Ours & \checkmark  & \checkmark  & 66.95  & 74.72  & 73.16  & 72.13  \\ \hline 
\multirow{4}{*}{30} & V0   & × & × & 53.74 & 66.94 & 65.86 & 62.01 \\
                    & V2   & × & \checkmark  & 69.26 & 66.67 & 61.67 & 66.04 \\
                    & V4   & \checkmark  & × & 80.00 & 56.10 & 80.79 & 68.35 \\
                    & Ours & \checkmark  & \checkmark  & 64.63  & 72.38  & 84.18  & 72.68 \\ \hline \hline 
\end{tabular} 
\end{table}

\renewcommand{\arraystretch}{1.6}
\begin{table*}[h]
\centering
\footnotesize
\caption{Ablation experiments: Recognition performance (\%) of different ablation configurations under different numbers of training samples in FUSAR-Ship. EFAA and DHFR stand for embedded feature augmenter with adaptive loss and dynamic hierarchical-feature refiner, respectively.}
\label{ablation_tab3}
\begin{tabular}{c|c|cc|cccccc}
\hline \hline 
\multirow{2}{*}{\begin{tabular}[c]{@{}c@{}}Training\\ Number\end{tabular}} &
  \multirow{2}{*}{Method} &
  \multirow{2}{*}{EFAA} &
  \multirow{2}{*}{DHFR} &
  \multirow{2}{*}{Bulk Carrier} &
  \multirow{2}{*}{Cargo Ship} &
  \multirow{2}{*}{Fishing} &
  \multirow{2}{*}{Other type of ship} &
  \multirow{2}{*}{Tanker} &
  \multirow{2}{*}{Average} \\
                    &      &   &   &       &       &       &       &       &       \\ \hline 
\multirow{4}{*}{20} & V0   & × & × & 86.96 & 42.86 & 46.47 & 61.67 & 60.94 & 52.97 \\
                    & V2   & × & \checkmark & 87.35 & 43.94 & 49.48 & 67.41 & 62.50 & 55.45 \\
                    & V4   & \checkmark & × & 86.17 & 41.06 & 58.17 & 73.76 & 41.67 & 56.95 \\
                    & Ours & \checkmark & \checkmark & 83.95  & 52.38  & 58.54  & 67.22  & 62.71  & 60.86  \\ \hline 
\multirow{4}{*}{30} & V0   & × & × & 84.36 & 45.15 & 55.09 & 63.33 & 71.19 & 55.74 \\
                    & V2   & × & \checkmark & 84.77 & 44.92 & 57.19 & 61.81 & 69.49 & 55.95 \\
                    & V4   & \checkmark & × & 86.01 & 55.30 & 54.04 & 67.61 & 35.78 & 59.71 \\
                    & Ours & \checkmark & \checkmark & 84.58  & 49.79  & 52.81  & 74.54  & 60.94  & 61.55  \\ \hline 
\multirow{4}{*}{40} & V0   & × & × & 78.54 & 49.97 & 60.00 & 70.23 & 82.41 & 60.73 \\
                    & V2   & × & \checkmark & 81.97 & 49.00 & 54.44 & 76.07 & 78.70 & 61.82 \\
                    & V4   & \checkmark & × & 81.12 & 46.83 & 66.17 & 75.19 & 79.63 & 62.52 \\
                    & Ours & \checkmark & \checkmark & 81.55  & 56.38  & 60.13  & 77.73  & 75.00  & 66.63 \\ \hline \hline 
\end{tabular} 
\end{table*}

\renewcommand{\arraystretch}{1.5}
\begin{table*}[]
\centering
\caption{Recognition Performances of 10 classes under different training datasets in MSTAR}
\label{mstar-results}
\setlength{\tabcolsep}{2.1mm}{
\begin{tabular}{c|ccccccccccccc}
\hline \hline 
\multirow{2}{*}{Class} & \multicolumn{13}{c}{Training Number in Each Class}                                                                              \\ \cline{2-14} 
                       & 5       & 10      & 20      & 25      & 30      & 40      & 55      & 60      & 80      & 100     & 110     & 165     & 220     \\ \hline 
BMP2 & 33.33\% & 70.77\% & 90.77\% & 86.67\% & 93.85\% & 97.44\% & 96.41\% & 98.97\% & 98.97\% & 98.97\% & 97.95\% & 98.97\% & 99.74\% \\
BRDM2 & 81.39\% & 72.99\% & 98.54\% & 97.08\% & 94.16\% & 96.35\% & 98.54\% & 98.18\% & 99.64\% & 98.54\% & 98.91\% & 99.82\% & 99.82\% \\
BTR60 & 82.56\% & 83.59\% & 94.36\% & 89.74\% & 98.46\% & 96.92\% & 97.44\% & 97.44\% & 97.44\% & 98.97\% & 96.92\% & 96.92\% & 97.95\% \\
BTR70 & 47.96\% & 83.67\% & 94.90\% & 93.37\% & 97.96\% & 96.43\% & 99.49\% & 99.49\% & 98.47\% & 99.49\% & 99.49\% & 99.49\% & 99.49\% \\
D7 & 90.51\% & 97.45\% & 94.16\% & 99.64\% & 99.64\% & 99.64\% & 99.64\% & 99.64\% & 99.64\% & 99.64\% & 99.96\% & 99.82\% & 99.82\% \\
2S1 & 73.72\% & 87.23\% & 94.89\% & 98.54\% & 97.81\% & 99.27\% & 99.27\% & 98.54\% & 99.64\% & 99.64\% & 99.96\% & 99.82\% & 99.82\% \\
T62 & 80.22\% & 85.35\% & 99.27\% & 99.63\% & 98.90\% & 98.53\% & 99.63\% & 99.63\% & 99.63\% & 99.63\% & 99.96\% & 99.82\% & 99.82\% \\
T72 & 62.24\% & 97.96\% & 99.49\% & 99.49\% & 99.49\% & 99.49\% & 99.49\% & 99.49\% & 99.49\% & 99.49\% & 99.95\% & 99.74\% & 99.74\% \\
ZIL131 & 86.50\% & 94.53\% & 99.64\% & 99.64\% & 98.91\% & 98.54\% & 98.91\% & 99.64\% & 99.64\% & 99.64\% & 99.96\% & 99.82\% & 99.82\% \\
ZSU235 & 93.43\% & 98.18\% & 99.64\% & 99.64\% & 99.64\% & 99.64\% & 99.64\% & 99.64\% & 99.64\% & 99.64\% & 99.96\% & 99.82\% & 99.82\% \\ \hline 
Average & 75.34\% & 87.59\% & 96.78\% & 96.87\% & 97.94\% & 98.31\% & 98.93\% & 99.09\% & 99.30\% & 99.38\% & 99.40\% & 99.48\% & 99.63\%   \\ \hline \hline 
\end{tabular} }
\end{table*}

\begin{table}[tb]\color{blue}
\renewcommand{\arraystretch}{1.2}
\setlength\tabcolsep{4.5pt}
\caption{\blue{Training and Testing dataset under EOCs}}
\centering
\footnotesize 
\label{ttnumEOCs}
\begin{tabular}{lcclcc}
\toprule\toprule
Train      & Number & \begin{tabular}[c]{@{}c@{}}Depression\\ Angle\end{tabular}  & Test(EOC-D) & Number & \begin{tabular}[c]{@{}c@{}}Depression\\ Angle\end{tabular} \\ \hline
2S1        & 299    & \multirow{4}{*}{$\text{17}{}^\circ$}    & 2S1-b01    & 288 & \multirow{4}{*}{$\text{30}{}^\circ$}   \\ 
BRDM2      & 298    &   & BRDM2-E71  & 287 &    \\ 
T72        & 232    &   & T72-A64    & 288 &    \\ 
ZSU234     & 299    &   & ZSU234-d08 & 288 &    \\ \midrule \midrule
Train      & Number & \begin{tabular}[c]{@{}c@{}}Depression\\ Angle\end{tabular}  & Test(EOC-C) & Number & \begin{tabular}[c]{@{}c@{}}Depression\\ Angle\end{tabular} \\ \hline
BMP2       & 233    & \multirow{5}{*}{$\text{17}{}^\circ$}    & T72-S7     & 419 & \multirow{5}{*}{$\text{15}{}^\circ$, $\text{17}{}^\circ$}    \\ 
BRDM2      & 298    &   & T72-A32    & 572 &    \\ 
BTR70      & 233    &   & T72-A62    & 573 &    \\ 
T72        & 232    &   & T72-A63    & 573 &    \\ 
           &        &   & T72-A64    & 573 &    \\ \midrule \midrule
Train      & Number & \begin{tabular}[c]{@{}c@{}}Depression\\ Angle\end{tabular}  & Test(EOC-V) & Number & \begin{tabular}[c]{@{}c@{}}Depression\\ Angle\end{tabular} \\ \hline
BMP2       & 233    & \multirow{7}{*}{$\text{17}{}^\circ$}    & T72-SN812  & 426 & \multirow{7}{*}{$\text{15}{}^\circ$, $\text{17}{}^\circ$}    \\ 
           &        &   & T72-A04    & 573 &    \\ 
BRDM2      & 298    &   & T72-A05    & 573 &    \\ 
           &        &   & T72-A07    & 573 &    \\ 
BTR70      & 233    &   & T72-A10    & 567 &    \\ 
           &        &   & BMP2-9566  & 428 &    \\ 
T72        & 232    &   & BMP2-C21   & 429 &    \\ \bottomrule\bottomrule
\end{tabular}
\end{table}

\begin{table}[tb]\color{blue}
\renewcommand{\arraystretch}{1.2}
\setlength\tabcolsep{7.1pt}
\centering
\footnotesize
\caption{\blue{Recognition Performance (\%) under EOCs on MSTAR}}
\label{EOCs-perf}
\begin{tabular}{ccccccc}
\toprule\toprule
\multicolumn{7}{c}{EOC-D}                                                                                                    \\ \midrule
\multirow{2}{*}{Class} & \multicolumn{6}{c}{Training Number in Each Class}                                                   \\ \cline{2-7}
                       & 5              & 10             & 20             & 40             & 60             & 100            \\ \hline 
BRDM2-E71              & 61.89          & 97.55          & 97.90          & 86.71          & 98.60          & 98.95          \\
2S1-b01                & 66.32          & 92.71          & 97.57          & 98.96          & 99.31          & 99.65          \\
T72-132                & 70.14          & 34.03          & 67.36          & 86.46          & 81.25          & 84.72          \\
ZSU234-d08             & 84.03          & 89.93          & 96.88          & 94.10          & 94.44          & 97.92          \\
Average                & \textbf{70.61} & \textbf{78.52} & \textbf{89.91} & \textbf{91.57} & \textbf{93.39} & \textbf{95.30} \\ \midrule
\multicolumn{7}{c}{EOC-C}                                                                                                    \\ \midrule
\multirow{2}{*}{Class} & \multicolumn{6}{c}{Training Number in Each Class}                                                   \\ \cline{2-7}
                       & 5              & 10             & 20             & 40             & 60             & 100            \\ \hline
T72-A32                & 81.26          & 86.87          & 94.75          & 97.20          & 92.99          & 96.85          \\
T72-A62                & 81.15          & 88.31          & 94.24          & 95.29          & 95.81          & 97.91          \\
T72-A63                & 72.60          & 80.28          & 88.13          & 90.05          & 88.31          & 92.32          \\
T72-A64                & 77.84          & 87.26          & 94.07          & 91.62          & 97.38          & 97.73          \\
T72-S7                 & 96.90          & 94.27          & 99.28          & 97.85          & 99.05          & 99.76          \\
Average                & \textbf{81.10} & \textbf{87.01} & \textbf{93.80} & \textbf{94.20} & \textbf{94.46} & \textbf{96.75} \\ \midrule
\multicolumn{7}{c}{EOC-V}                                                                                                    \\ \midrule
\multirow{2}{*}{Class} & \multicolumn{6}{c}{Training Number in Each Class}                                                   \\ \cline{2-7}
                       & 5              & 10             & 20             & 40             & 60             & 100            \\ \hline
BMP2-9566              & 67.76          & 83.18          & 83.88          & 92.99          & 96.73          & 96.50          \\
BMP2-C21               & 70.86          & 82.28          & 84.85          & 94.17          & 97.20          & 99.30          \\
T72-SN812              & 95.77          & 87.32          & 95.07          & 98.36          & 99.30          & 99.53          \\
T72-A04                & 80.45          & 88.13          & 96.16          & 95.46          & 97.38          & 98.25          \\
T72-A05                & 86.91          & 90.40          & 97.91          & 98.08          & 99.13          & 98.25          \\
T72-A07                & 78.71          & 90.94          & 96.68          & 97.56          & 99.30          & 97.91          \\
T72-A10                & 72.84          & 91.01          & 89.07          & 92.42          & 98.41          & 98.59          \\
Average                & \textbf{79.15} & \textbf{88.00} & \textbf{92.43} & \textbf{95.63} & \textbf{98.26} & \textbf{98.32} \\ \bottomrule\bottomrule
\end{tabular}
\end{table}

\subsection{Ablation Experiments}
In this subsection, we discuss the removal of the main innovations separately to run recognition experiments of different configurations with the MSTAR, OpenSARShip and FUSAR-Ship datasets. The comparison of the recognition performances of these configurations is followed, consisting of three different numbers of training samples for each class—100, 60, and 20 for MSTAR, 30, 20 and 10 for OpenSARShip, and 40, 30 and 20 for FUSAR-Ship. For MSTAR, the training images and testing images are taken at depression angles of $17^{\circ}$ and $15^{\circ}$, respectively, following the general settings of SAR ATR \cite{MIT}. 
Table \ref{ttnumMSTAR} summarizes the experimental setup for the training and testing datasets. During the experiments, a limited set of training samples is randomly chosen from the original training sample set shown in Table \ref{ttnumMSTAR}.

There are {a total of} five different configurations {involved in our method,} as shown in Table \ref{ablation_tab}. The first one, V0, {represents} our method without the three innovations---the embedded feature augmenter, dynamic hierarchical-feature refiner, and adaptive loss. The second one, V1, is our method with the embedded feature augmenter, but without the dynamic hierarchical-feature refiner {and} adaptive loss. The third one, V2,
is our method with the embedded feature augmenter {and adaptive loss} but without the dynamic hierarchical-feature refiner. The fourth one, V3, {is our method without just the adaptive loss. The last one, V4, is our method just without the embedded feature augmenter and adaptive loss.} We also {present} the recognition {performance} of the full version of our method.

\subsubsection{Comparisons with Different Training Samples}
From the comparisons of the four different configurations and the full version of our method with three different numbers of training samples, there are two {prevailing characteristics}. 1) {Our full method is} more effective facing fewer training samples. For example, the recognition ratios of V4 and the full version with 100 samples {in} each class are 88.91\% and 99.38\%, respectively, and {those} with 20 samples {in} each class are 80.08\% and 96.78\%, respectively. 2) {The embedded feature augmenter and adaptive loss significantly improve the recognition performance for the three different training sample sets, shown by comparing the recognition of V0, V1, and V4, with the full method. The similar situations can be also observed by the ablation experiments under OpenSARShip and FUSAR-Ship dataset.}

\subsubsection{Comparisons with Different Configurations}
The comparisons of the four different configurations and full version have illustrated the following. 1) The embedded feature augmenter can boost recognition performance with a limited number of training samples, revealed by comparing the recognition of V0, V1, and V3. Recognition using the limited training set can be enhanced by roughly searching and augmenting the hard-to-recognize samples. 2) The dynamic hierarchical-feature refiner can play a more effective role when combined with the embedded feature augmenter, as revealed by comparing the recognition ratios of V1 and V2. We suggest that this is because the embedded feature augmenter helps the dynamic hierarchical-feature refiner to capture the dynamic nature of the SAR images by augmenting the hard-to-recognize samples. 3) The adaptive loss can further improve recognition performance owing to the embedded feature augmenter, which is found by comparing the recognition of V2, V3, and the full version. The subtle control of the intensities of clustering or distancing from the adaptive loss is useful for recognition with a limited training sample set.

The comparisons of the ablation experiments above {show} that our main innovations {effectively improve} recognition performance {when handling limited numbers of training samples}, and they can also {cooperate to further tackle subproblems encountered in this scenario}. In conclusion, the effectiveness and {robustness} of our method have been validated. {The feature distribution of different configurations are also shown in Fig. \ref{ablation_fig}.}

\subsection{Recognition Performances Under MSTAR, OpenSARShip, and FUSAR-Ship}
In this subsection, the recognition performances under three datasets are presented as {follows}. In each subsubsection, the training and testing data of the dataset, {recognition results under the datasets, and analyses of the recognition performance are presented}.

\subsubsection{Recognition Results under MSTAR}

\blue{The effectiveness of our method is validated by the recognition of the standard operating conditions (SOC) and extended operating conditions (EOCs). 
For SOC, the training and testing images are taken at depression angles of $17^{\circ}$ and $15^{\circ}$, respectively. 
During the experiments of SOC, the limited training sample set is randomly chosen from the original training sample set from Table \ref{ttnumMSTAR}.
EOCs consisted of EOC-C (configuration variant), EOC-D (depression variant), and EOC-V (version variant). The experiments on the EOCs are similar to practical situations, that is, it is difficult to achieve high performance because of the large variance between the training and testing samples. The training sample set for EOCs is randomly chosen from Table \ref{ttnumEOCs}.}

Table \ref{mstar-results} shows the recognition results {in the form of a confusion matrix for numbers of the training samples for each class ranging from 5 to 220. For the training samples of} 100 and 80 for each class, the recognition ratios of the ten classes {reach} 99.38\% and 99.30\%, respectively. {For the training samples of 55 and 40 for each class, the recognition ratios are 98.93\% and 98.31\%, respectively. The recognition ratios for 25 and 20 in each class reach 96.87\% and 96.78\%, respectively.} 
From the results, {our method achieves high recognition performance for sufficient numbers of training samples}. When the training samples are reduced to 20 {and} 25 {for} each class, our method still obtains {recognition ratios greater than 96.78\%}. 
{This stable recognition performance for limited training sample sets is useful and significant for the application of SAR ATR methods.}

\blue{Table \ref{EOCs-perf} shows the recognition results of EOCs when the training sample for each class is ranging from 5 to 100. For the training samples of 100 each class, the recognition ratios of EOCs achieved state-of-the-art. When the training samples for each class are 60 or 40, the recognition ratios of EOCs stay above 91.00\%, which indicates the proposed method faces a small data pressure in practical application of SAR ATR.
when the training samples for each class is decreased to 5, 10 and 20, the recognition ratios of EOCs is gradually reduced, and with 5 SAR images each class, the recognition ratios of EOCs stay within the range of 70.00\% to 82.00\%. }

\blue{The recognition performance of the proposed method under SOC and EOCs constraints with limited training samples in the MSTAR dataset suggests its robustness and effectiveness, maintaining a high level of accuracy.}

\renewcommand{\arraystretch}{1.5}
\begin{table}[]
\centering
\caption{Number of images and imaging conditions of different targets in OpenSARShip}
\label{opensarset}
\setlength\tabcolsep{2pt}
\begin{tabular}{c|c|ccc}
\hline \hline 
Class          & Imaging Condition                                                                  & \begin{tabular}[c]{@{}c@{}}Training\\ Number\end{tabular} & \begin{tabular}[c]{@{}c@{}}Testing\\ Number\end{tabular} & \begin{tabular}[c]{@{}c@{}}Total\\ Number\end{tabular} \\ \hline 
Bulk Carrier   & \multirow{6}{*}{\begin{tabular}[c]{@{}c@{}} VH and VV, C ban\\  Resolution=$5-20$m\\ Incident angle=$20^{\circ}-45^{\circ}$ \\ Elevation sweep angle=$\pm 11^{\circ}$\\ ${\text{Rg20}}m \times {\text{az}}22m$\end{tabular}} & 200                                                       & 475                                                      & 675                                                    \\ \cline{1-1} \cline{3-5}
Container Ship &                                                                                    & 200                                                       & 811                                                      & 1011                                                   \\ \cline{1-1} \cline{3-5}
Tanker         &                                                                                    & 200                                                       & 354                                                      & 554                                                    \\ \cline{1-1} \cline{3-5}
Cargo          &                                                                                    & 200                                                       & 557                                                      & 757                                                    \\ \cline{1-1} \cline{3-5}
Fishing        &                                                                                    & 200                                                       & 121                                                      & 321                                                    \\ \cline{1-1} \cline{3-5}
General Cargo  &                                                                                    & 200                                                       & 165                                                      & 365     \\ \hline \hline                                               
\end{tabular}
\end{table}

\renewcommand{\arraystretch}{1.5}
\begin{table*}[]
\centering
\caption{Recognition performances of three classes under different training datasets in OpenSARShip}
\label{opensarship3-r}
\setlength{\tabcolsep}{1.3mm}{
\begin{tabular}{c|ccccccccc}
\hline \hline 
\multirow{2}{*}{Class} & \multicolumn{9}{c}{Training Number in Each Class}                                       \\ \cline{2-10} 
                       & 10      & 20      & 30      & 40      & 60      & 70      & 80      & 100     & 200     \\ \hline 
  Bulk Carrier           & 48.84\% & 66.95\% & 64.63\% & 59.79\% & 75.37\% & 75.16\% & 74.74\% & 75.58\% & 75.14\% \\
Container Ship         & 77.19\% & 74.72\% & 72.38\% & 83.97\% & 77.68\% & 79.65\% & 81.50\% & 85.70\% & 89.92\% \\
Tanker                 & 71.47\% & 73.16\% & 84.18\% & 79.66\% & 82.77\% & 83.62\% & 85.88\% & 83.62\% & 91.08\% \\ \hline 
Average                & 67.74\% & 72.13\% & 72.68\% & 76.04\% & 78.11\% & 79.21\% & 80.49\% & 82.32\% & 85.40\% \\ \hline \hline 
\end{tabular} }
\end{table*}

\renewcommand{\arraystretch}{1.5}
\begin{table*}[]
\centering
\caption{Recognition performances of six classes under different training datasets in OpenSARShip}
\label{opensarship6-r}
\setlength{\tabcolsep}{1.3mm}{
\begin{tabular}{c|ccccccccc}
\hline \hline 
\multirow{2}{*}{Class} & \multicolumn{9}{c}{Training Number in Each Class}                                       \\ \cline{2-10} 
                       & 10      & 20      & 30      & 40      & 60      & 70      & 80      & 100     & 200     \\ \hline 
Bulk Carrier           & 48.00\% & 66.53\% & 53.26\% & 61.05\% & 66.53\% & 61.26\% & 60.21\% & 68.42\% & 63.29\% \\
Container Ship         & 72.26\% & 59.56\% & 69.42\% & 74.48\% & 71.15\% & 79.16\% & 78.18\% & 81.01\% & 88.46\% \\
Tanker                 & 41.53\% & 51.41\% & 52.82\% & 41.24\% & 50.28\% & 52.54\% & 53.11\% & 53.95\% & 69.52\% \\
Cargo                  & 31.06\% & 39.32\% & 39.86\% & 36.27\% & 40.57\% & 37.52\% & 43.63\% & 42.19\% & 62.19\% \\
Fishing                & 85.12\% & 81.82\% & 86.78\% & 83.47\% & 93.39\% & 92.56\% & 85.95\% & 89.26\% & 84.48\% \\
General Cargo          & 18.18\% & 33.94\% & 44.24\% & 37.58\% & 47.27\% & 45.45\% & 45.45\% & 46.06\% & 27.98\% \\ \hline 
Average                & 51.03\% & 54.57\% & 56.50\% & 56.58\% & 59.93\% & 61.01\% & 61.62\% & 64.12\% & 67.03\%  \\ \hline \hline 
\end{tabular} }
\end{table*}

\subsubsection{Recognition Results under OpenSARShip}

The OpenSARShip dataset contains several ship classes {comprising} the most common and representative ships, occupying 90\% of the international shipping market \cite{open4}.
Following \cite{open2,open3,open4}, {two experiments, under three different classes and six classes, are conducted.} As shown in Table \ref{opensarset}, the experiments {under three classes involve} bulk carriers, container ships, and tanks. {The six-class experiments include cargo ships, fishing, and general cargo along with the previous three classes.} The preprocessing method of the training and testing images is the same as {that under the MSTAR dataset}.

The recognition ratios are shown in Table \ref{opensarship3-r} and Table \ref{opensarship6-r} {for} training samples for each class {ranging} from 10 to 200. 
The range of the training number is set according to the {corresponding} methods \cite{open2,open3,open4}. 

From the results, it is clear that our method can achieve superior recognition performance for the SAR ship images in both {groups of three classes and six classes}. {For the recognition of three classes, our method shows robustness for training samples ranging from 40 to 10. When the training samples in each class decrease from 40 to 20, the recognition ratios drop by 3.91\%, from 76.04\% to 72.13\%. The recognition ratios fall by 4.45\% from 80.49\% to 76.04\% for training samples in each class decreasing from 80 to 40. For a decrease from 200 to 100 training samples, the recognition ratios decrease by 3.08\%, from 85.40\% to 82.32\%. A reduction in the training samples by half demonstrates the robustness of our method facing fewer training samples. Robustness of recognition performance under a limited training sample set is a good property for the application of SAR ATR methods. The same phenomenon is also shown in the recognition of six classes.}

\renewcommand{\arraystretch}{1.5}
\begin{table}[]
\centering
\caption{Number of images and imaging conditions of different targets in FUSAR-Ship}
\label{fusarset}
\setlength\tabcolsep{0.2pt}
\begin{tabular}{c|c|ccc}
\hline \hline 
Class          & Imaging Condition                                                                  & \begin{tabular}[c]{@{}c@{}}Training\\ Number\end{tabular} & \begin{tabular}[c]{@{}c@{}}Testing\\ Number\end{tabular} & \begin{tabular}[c]{@{}c@{}}Total\\ Number\end{tabular} \\ \hline 
Bulk Carrier          & \multirow{5}{*}{\begin{tabular}[c]{@{}c@{}}VH and VV, C band\\  Resolution $=0.5-500$m\\ Incident angle $=10^{\circ}-60^{\circ}$ \\ Elevation sweep angle $=\pm 20^{\circ}$\\ ${\text{Rg20}}m \times {\text{az}}22m$\end{tabular}} & 100                  & 173                  & 273                  \\ \cline{1-1} \cline{3-5}
Cargo Ship            &                                                                                    & 100                  & 1593                 & 1693                 \\ \cline{1-1} \cline{3-5}
Fishing              &                                                                                    & 100                  & 685                  & 785                  \\ \cline{1-1} \cline{3-5}
Other Types of ship      &                                                                                    & 100                  & 1507                 & 1607                 \\ \cline{1-1} \cline{3-5}
Tanker               &                                                                                    & 100                  & 48                   & 148                  \\ \hline \hline                                               
\end{tabular} 
\end{table}

\renewcommand{\arraystretch}{1.5}
\begin{table}[]
\centering
\caption{Recognition performances of {five classes under different training datasets} in FuSAR-Ship}
\label{fusar-r}
\setlength{\tabcolsep}{0.9mm}{
\begin{tabular}{c|cccccc}
\hline \hline 
\multirow{2}{*}{Class} & \multicolumn{6}{c}{Labeled Number in Each   Class}        \\ \cline{2-7} 
                       & 20      & 30      & 40      & 60      & 80      & 100     \\ \hline
Bulk Carrier            & 83.95\% & 84.58\% & 81.55\% & 88.26\% & 82.38\% & 90.75\% \\
Cargo Ship              & 52.38\% & 49.79\% & 56.38\% & 69.32\% & 61.00\% & 62.21\% \\
Fishing                & 58.54\% & 52.81\% & 60.13\% & 33.24\% & 64.68\% & 64.53\% \\
Other Types of ship     & 67.22\% & 74.54\% & 77.73\% & 79.90\% & 78.19\% & 77.97\% \\
Tanker                 & 62.71\% & 60.94\% & 75.00\% & 69.32\% & 83.82\% & 81.25\% \\ \hline 
Average                & 60.86\% & 61.55\% & 66.63\% & 67.95\% & 69.41\% & 70.00\% \\ \hline \hline 
\end{tabular} }
\end{table}

For the recognition of SAR ship images, our method can achieve superior performances {under different numbers of classes}. The experiments on the OpenSARship dataset also {validate} the effectiveness and robustness of our method {for} different imaging scenes, including land, sea, and offshore. 


\begin{table*}[]
\renewcommand{\arraystretch}{1.5}
\setlength\tabcolsep{3pt}
\centering
\caption{Comparison of Performances (\%) in MSTAR}
\label{comparisonMSTAR1}
\begin{tabular}{c|c|cccccccc}
\hline\hline
\multicolumn{2}{c|}{\multirow{2}{*}{Algorithms}} & \multicolumn{8}{c}{Image Number for Each Class}                                                         \\ \cline{3-10} 
\multicolumn{2}{c|}{}                            & \multicolumn{1}{c}{20} & \multicolumn{1}{c}{40} & \multicolumn{1}{c}{55} &\multicolumn{1}{c}{80} &\multicolumn{1}{c}{110} &\multicolumn{1}{c}{165} &\multicolumn{1}{c}{220} & \multicolumn{1}{c}{All data} \\ \hline
\multirow{6}{*}{Traditional} &PCA+SVM \cite{comparison1}               & 76.43             & 87.95    &   -      & 92.48        &- &- &-     & 94.32                 \\
&ADaboost \cite{comparison1}              & 75.68             & 86.45     & -       & 91.45   &- &- &-          & 93.51                 \\
&DGM \cite{comparison1}                   & 81.11             & 88.14        & -    & 92.85    &- &- &-         & 96.07                 \\
&LNP \cite{addc1}   &- & -  & 92.04  &   -      & 94.11  & 95.97      & 96.05         &  -        \\
&PSS-SVM \cite{addc2}          &- &  -              & 95.01        & -  & 95.67           & 96.02          & 96.11  & -         \\ \hline 
\multirow{6}{*}{\begin{tabular}[c]{@{}c@{}}Data\\augmentation-based\end{tabular}} &GAN-CNN1 \cite{comparison1}                  & 81.80             & 88.35      & -      & 93.88    &- &- &-         & 97.03                 \\
& GAN-CNN2 \cite{comparison1}               & 84.39             & 90.13    &   -      & 94.91   &- & -& -         & 97.53                 \\ 
&MGAN-CNN \cite{comparison1}              & 85.23             & 90.82    & -        & 94.91   & -&- & -         & 97.81                 \\ 
&Triple-GAN\cite{addc3}       &-&-               & 95.70      &-     & 95.97           & 96.13                   & 96.46      &-        \\
&Improved-GAN     \cite{addc4}    &-&-             & 87.52        &-   & 95.02           & 97.26                     & 98.07    &-       \\
&Semi-supervised GAN\cite{addc5}    &-&-            & 95.72      &-     & 97.22           & 97.97            & 98.14         &-    \\ 
&Semi-Supervised \cite{open4} & 98.23 & 99.85 & 99.81 & 99.85 & - & - &-  &-  \\\hline 
\multirow{4}{*}{\begin{tabular}[c]{@{}c@{}}Novel\\model/module\end{tabular}} 
&Deep CNN \cite{comparison2}                  & 77.86             & 86.98     &     -   & 93.04    &- &- &-         & 95.54                 \\
&Improved DNN \cite{comparison3}                  & 79.39             & 87.73    &  -       & 93.76    &- &- &  -       & 96.50                 \\
&Simple CNN \cite{compared2}                  & 75.88             & -     &      -      & -   &- &- &  -            & -                     \\
&Metric learning\cite{compared2}            & 82.29             & -      &    -       & -      & -&- &    -       & -                     \\ \hline 
\multicolumn{2}{c|}{Ours   }                                  & 96.78    & 98.31 & 98.93  & 99.30  &99.40 &99.48 & 99.63 & -                     \\
\hline\hline
\end{tabular}
\end{table*}

\subsubsection{Recognition Results under FUSAR-Ship}

As shown above,  {recognition on the FUSAR-Ship dataset} is more complex than {that on the} OpenSARship dataset. We choose five ship classes {from the list for} the OpenSARship dataset, and one type {has been} named {as other types of ship}. The first five ship classes are the most common in the world shipping market. The last ship class, other types of ships, is a set that includes most of the other ships {apart from} the first five ship classes. Thus, this class has more overlap than the other ship classes, {requiring} more robustness and efficiency {in} the method. Table \ref{fusarset} shows the original dataset of the training and testing sets in {the} FUSAR-Ship dataset. 

The recognition performances of the five classes on the FUSAR-Ship dataset are shown in Table \ref{fusar-r}. The number of training samples ranges from 100 to 20. From the results, the recognition performances of our method are superior. When the training samples are decreased from 100 to 60, the recognition ratios fall by 2.05\%, from 70.00\% to 67.95\%. For a decrease in the number of training samples from 40 to 20, the recognition ratios decrease by 5.77\%, from 66.63\% to 60.86\%. The larger drop (5.08\%) between the recognition ratios of 40 and 30 per class may be caused by the complexity of the FUSAR-Ship dataset.

From the recognition performances on {the} MSTAR, OpenSARship, and FUSAR-Ship datasets, the effectiveness and robustness of our method are validated {under the condition of} limited training samples. {Furthermore}, it is clear that our method can handle different imaging scenes, target classes (vehicle and ship), and complex imaging conditions.

\subsection{Comparison}
In this subsection, the performances of our method are compared with other state-of-the-art methods for SAR ATR under different limited training sample datasets. We compare our method with two types of methods—methods based on data augmentation and methods constructing novel models or modules.

\subsubsection{Comparison under MSTAR}
In this section, {a} comparison with state-of-the-art methods on MSTAR is presented in Table \ref{comparisonMSTAR1}. {These methods comprise three main types}: traditional methods, data augmentation-based methods, and methods constructing novel models or modules. The introduction of each method is given below.

Traditional methods, including PCA+SVM, ADaboost, DGM, LNP, {and} PSS-SVM, are compared with our method. 
Then, data augmentation-based methods, including GAN-CNN1, GAN-CNN2, MGAN-CNN, triple GAN, improved GAN, semi-supervised GAN and Semi-Supervised are also gathered for comparison. It should be noted that Semi-Supervised used $k$ samples each class as labeled samples, and the resting samples of each class as unlabeled samples.
Furthermore, state-of-the-art deep-learning-based methods proposing novel models or modules are also selected, including deep CNN, improved DNN, simple CNN, {and} metric learning.

From the comparison in Table \ref{comparisonMSTAR1}, it is clear that our method has achieved {state-of-the-art recognition performance with a wide range of training samples}. Our method achieves the best performance without {the use of} additional datasets or semi-supervised architecture.

\renewcommand{\arraystretch}{2}
\begin{table}[]
\centering
\caption{Comparison of Performances (\%) of three classes under OpenSARShip (The number in parentheses is the number of the training samples for each method)}
\label{3comparisonOPEN}
\begin{tabular}{c|ccc}
\hline\hline
\multirow{2}{*}{Methods} &
  \multicolumn{3}{c}{Number range of training images in each class} \\ \cline{2-4} 
 &
  \multicolumn{1}{c|}{1 to 50} &
  \multicolumn{1}{c|}{51 to 100} &
  101 to 338 \\ \hline
Semi-Supervised \cite{open4} &
  \multicolumn{1}{c|}{\begin{tabular}[c]{@{}c@{}}61.88 (20)\\ 64.73 (40)\end{tabular}} &
  \multicolumn{1}{c|}{68.67 (80)} &
  \begin{tabular}[c]{@{}c@{}}71.29 (120)\\ 74.96 (240)\end{tabular} \\ \hline
Supervised \cite{open4} &
  \multicolumn{1}{c|}{\begin{tabular}[c]{@{}c@{}}58.24 (20)\\ 62.09 (40)\end{tabular}} &
  \multicolumn{1}{c|}{65.63 (80)} &
  \begin{tabular}[c]{@{}c@{}}68.75 (120)\\ 70.83 (240)\end{tabular} \\ \hline
CNN\cite{compared2} &
  \multicolumn{1}{c|}{62.75 (50)} &
  \multicolumn{1}{c|}{68.52 (100)} &
  73.68 (200) \\ \hline
CNN+Matrix\cite{compared2} &
  \multicolumn{1}{c|}{72.86 (50)} &
  \multicolumn{1}{c|}{75.31 (100)} &
  77.22 (200) \\ \hline 
PFGFE-Net\cite{compared3} &
  \multicolumn{1}{c|}{-} &
  \multicolumn{1}{c|}{-} &
  79.84 (338) \\ \hline
MetaBoost\cite{compared4} &
  \multicolumn{1}{c|}{-} &
  \multicolumn{1}{c|}{-} &
  80.81 (338) \\ \hline
Proposed &
  \multicolumn{1}{c|}{\begin{tabular}[c]{@{}c@{}}72.13 (20)\\ 76.04 (40)\end{tabular}} &
  \multicolumn{1}{c|}{\begin{tabular}[c]{@{}c@{}}80.49 (80)\\ 82.32 (100)\end{tabular}} &
  85.40 (200) \\ \hline\hline
\end{tabular}
\end{table}

\renewcommand{\arraystretch}{1.2}
\begin{table*}[htb!]
\centering
\footnotesize
\caption{Comparison with effective deep learning networks under constant training samples of OpenSARShip (The training samples of other methods under 3 classes and 6 classes are evaluated under 338 and 200 samples in each class.)}
\label{3&6COMPARISONOpenSARShip}
\setlength\tabcolsep{3.9pt}
\begin{tabular}{c|c|cccccccc}
\hline \hline  
\multicolumn{2}{c|}{\multirow{2}{*}{Model}} & \multicolumn{4}{c}{3 classes}                                                                                                                                                                                                                                                 & \multicolumn{4}{c}{6 classes}                          \\ \cline{3-10} 
\multicolumn{2}{c|}{}    & Recall (\%)                                                       & Precision (\%)                                                    & F1 (\%)                                                           & Acc (\%)                                                          & Recall (\%) & Precision (\%) & F1 (\%)    & Acc (\%)   \\ \hline 
\multirow{15}{*}{\begin{tabular}[c]{@{}c@{}}Effective \\ deep learning \\ methods \end{tabular}} & LeNet-5 \cite{2compared1}                & 65.15±1.12 & 60.54±2.47 & 62.73±1.52 & 65.74±1.50 & 49.14±0.64 & 39.53±0.54 & 43.81±0.81 & 46.59±1.50 \\
& AlexNet \cite{2compared2}               & 68.51±3.04 & 65.52±1.23 & 66.94±1.51 & 70.22±0.68 & 53.40±0.61 & 44.39±0.60 & 48.48±0.87 & 50.98±2.14 \\
& VGG-11 \cite{2compared3} & 73.21±0.96 & 68.64±1.49 & 70.85±1.07 & 73.42±0.75 & 51.38±0.82 & 43.67±1.23 & 47.21±1.25 & 49.41±0.99 \\
& VGG-13 \cite{2compared3} & 72.59±1.29 & 67.24±1.75 & 68.79±0.92 & 73.03±0.86 & 51.32±0.38 & 43.06±1.68 & 46.83±0.90 & 49.70±1.36 \\
& GooLeNet \cite{2compared5} & 69.73±2.70 & 68.80±1.81 & 69.21±1.19 & 73.80±1.32 & 54.47±0.95 & 44.96±1.76 & 49.25±0.70 & 49.76±1.56 \\
& ResNet-18 \cite{2compared6} & 73.76±1.61 & 69.40±1.92 & 71.49±1.04 & 74.64±0.68 & 50.19±0.47 & 42.85±1.20 & 46.23±0.35 & 45.91±0.43 \\
& ResNet-34 \cite{2compared6} & 71.43±2.72 & 68.11±1.73 & 69.69±1.47 & 73.40±1.09 & 48.12±0.57 & 42.18±0.57 & 44.95±0.83 & 48.27±2.75 \\
& ResNet-50 \cite{2compared6} & 71.67±1.71 & 66.79±1.27 & 69.13±1.04 & 72.82±0.75 & 50.27±1.21 & 43.32±1.32 & 46.54±2.50 & 49.80±1.70 \\
& DenseNet-121 \cite{2compared10} & 72.55±3.88 & 69.56±2.17 & 70.93±1.60 & 74.65±0.68 & 55.51±1.30 & 46.52±1.48 & 50.62±0.74 & 53.49±1.47 \\
& DenseNet-161 \cite{2compared10} & 72.54±3.39 & 67.77±1.46 & 70.02±1.51 & 73.39±0.79 & 54.98±0.82 & 47.57±0.82 & 51.01±1.63 & 54.27±3.41 \\
& MobileNet-v3-Large \cite{2compared12}     & 65.12±2.53 & 60.75±1.72 & 62.84±1.73 & 66.13±0.92 & 49.95±0.58 & 42.14±0.62 & 45.71±0.65 & 46.60±2.61 \\
& MobileNet-v3-Small \cite{2compared12} & 67.23±1.59 & 61.85±1.69 & 64.42±1.41 & 66.71±0.87 & 48.28±0.75 & 40.75±0.73 & 44.20±0.57 & 44.41±1.10 \\
& SqueezeNet \cite{2compared13}    & 71.47±1.31 & 66.73±1.70 & 69.01±1.28 & 72.15±1.25 & 53.24±0.75 & 45.55±0.79 & 49.10±0.85 & 53.12±1.12 \\
& Inception-v4 \cite{2compared14}           & 69.26±3.16 & 67.43±2.39 & 68.28±1.97 & 72.44±0.70 & 54.92±0.69 & 46.46±0.49 & 50.34±1.31 & 54.55±3.52 \\
& Xception \cite{2compared15} & 71.56±3.00 & 68.60±1.67 & 70.00±1.29 & 73.74±0.86 & 52.21±0.94 & 44.03±1.15 & 47.77±1.11 & 49.56±1.47 \\ \hline 
\multirow{8}{*}{\begin{tabular}[c]{@{}c@{}}Methods for \\ SAR ATR \end{tabular}}  &  Wang et al. \cite{p4}            & 57.72±1.37 & 58.72±4.76 & 58.12±2.67 & 69.27±0.27 & 50.53±1.85 & 41.77±1.34 & 45.73±2.48 & 48.43±3.71 \\
& Hou et al. \cite{FUSAR}             & 69.33±2.00 & 69.44±2.42 & 66.76±1.64 & 67.41±1.13 & 48.76±0.79 & 41.22±0.74 & 44.67±1.21 & 47.44±2.01 \\
& Huang et al. \cite{2compared18}           & 74.74±1.60 & 69.56±2.38 & 72.04±1.60 & 74.98±1.46 & 54.09±0.81 & 47.58±1.66 & 50.63±1.79 & 54.78±2.08 \\
& Zhang et al. \cite{open3}           & 77.87±1.14 & 73.42±1.06 & 75.05±1.10 & 78.15±0.57 & 54.20±1.09 & 46.66±1.07 & 50.15±1.24 & 53.77±3.63 \\
& Zeng et al. \cite{2compared27}            & 74.99±1.55 & 74.05±1.75 & 74.52±1,02 & 77.41±1.74 & 55.66±1.23 & 47.16±1.70 & 50.96±1.18 & 55.26±2.36 \\
& Xiong et al. \cite{2compared21}          & 73.87±1.16 & 71.50±3.00 & 72.67±2.04 & 75.44±2.68 & 53.57±0.33 & 45.74±0.82 & 49.35±0.69 & 54.93±2.61 \\
& SF-LPN-DPFF \cite{2compared22}           & 78.83±1.32 & 76.45±1.16 & 77.62±1.23 & 79.25±0.83 & 54.49±0.70 & 48.61±1.32 & 51.38±1.26 & 56.66±1.54 \\ \hline 
\multicolumn{2}{c|}{Ours}           & \begin{tabular}[c]{@{}c@{}}\bf{85.38±1.09}\\ (200)\end{tabular} & \begin{tabular}[c]{@{}c@{}}\bf{85.76±1.21}\\ (200)\end{tabular} & \begin{tabular}[c]{@{}c@{}}\bf{85.57±1.13}\\ (200)\end{tabular} & \begin{tabular}[c]{@{}c@{}}\bf{85.40±0.59}\\ (200)\end{tabular}  & \begin{tabular}[c]{@{}c@{}}\bf{65.99±0.52}\\ (200)\end{tabular}        & \begin{tabular}[c]{@{}c@{}}\bf{68.69±1.10} \\ (200)\end{tabular}         & \begin{tabular}[c]{@{}c@{}}\bf{67.31±0.67}\\ (200)\end{tabular}       & \begin{tabular}[c]{@{}c@{}} \bf{67.03±0.41}\\ (200)\end{tabular}      \\ \hline \hline 
\end{tabular} 
\end{table*}

\subsubsection{Comparison under OpenSARship}
In this section, {a} comparison with state-of-the-art methods on {the} OpenSARship dataset is presented. For a more comprehensive and holistic comparison, we {execute} two types of comparison. {The first is a comparison of different training SAR ship samples,} as shown in Table \ref{3comparisonOPEN}. {The second is a comparison} with different effective deep learning networks under constant training SAR ship samples, as shown in Table \ref{3&6COMPARISONOpenSARShip}. The Recall, Precision, F1, and Acc {metrics} based on the confusion matrices are calculated, and the formulas are as follows:
\begin{equation}
{\rm{Recall}}=\frac{  \sum_{i=1}^{K}{\rm{recall}}_i} {K},{\rm{recall}}_i=\frac{TP_i}{TP_i+FN_i}  
\end{equation}
\begin{equation}
{\rm{Precision}}=\frac{ \sum_{i=1}^{K}{\rm{precision}}_i} {K} ,{\rm{precision}}_i=\frac{TP_i}{TP_i+FP_i} 
\end{equation}
\begin{equation}
{\rm{F1}}=\frac{ \sum_{i=1}^{K}{\rm{F1}}_i} {K},{\rm{F1}}_i=\frac{2\times {\rm{precision}}_i\times {\rm{recall}}_i}{{\rm{precision}}_i+{\rm{recall}}_i}  
\end{equation}
where $K$ denotes the number of classes. It should be noted, in Table \ref{3&6COMPARISONOpenSARShip}, that the {training} of other methods under 3 classes and 6 classes are evaluated {for} 338 and 200 samples {in} each class. {Our method uses less than} 200 samples for each class.
We follow the experimental setup of \cite{open2} and {display} some of the recognition results for comparison. For simplicity, we use the bands of the training sample number to show the recognition performance in Table \ref{3comparisonOPEN}. These methods can also be divided into three types, {shown in Table \ref{comparisonMSTAR1}.}

In Table \ref{3comparisonOPEN}, the results and comparison {show} that our method achieves {state-of-the-art recognition performance on the OpenSARship dataset for decreased numbers of training samples.}
In Table \ref{3&6COMPARISONOpenSARShip}, it is clear that our method outperforms other deep learning methods even if the training samples used by our method are {less than (three classes) or equals to (six classes) those of other methods.}

From the comparison under MSTAR and OpenSARship {datasets}, our method has achieved {state-of-the-art} recognition performance. The effectiveness and robustness of our method have been validated {from comparison under wide ranges} of training sample numbers.

\section{Conclusion}
The limited availability of training data in SAR images poses a significant challenge for ATR methods, hindering their practical application. Some existing methods for limited training data treat all samples equally, ignoring the dynamic nature of SAR images and thus limiting their recognition performance. 
In this study, we proposed a SAR ATR method that uses an embedded feature augmenter and dynamic hierarchical-feature refiner for use with limited training sample sets. Our method first applies an embedded feature augmenter, a technique that mines crucial training samples for recognition with limited data and mixes their features and labels in a targeted manner to enhance the recognition ability of the network. A dynamic hierarchical-feature refiner is also implemented, which generates dynamic kernels to adapt to the dynamic nature of SAR images and obtain more robust overall features. 
Finally, a classifier is used to achieve accurate recognition with limited training data. Experimental results on the MSTAR, OpenSARship, and FUSAR datasets demonstrate that our method greatly improves recognition performance under such limitations.

\bibliographystyle{IEEEtran}
\bibliography{references}

\newpage

\begin{IEEEbiography}[{\includegraphics[width=1in,height=1.25in,clip,keepaspectratio]{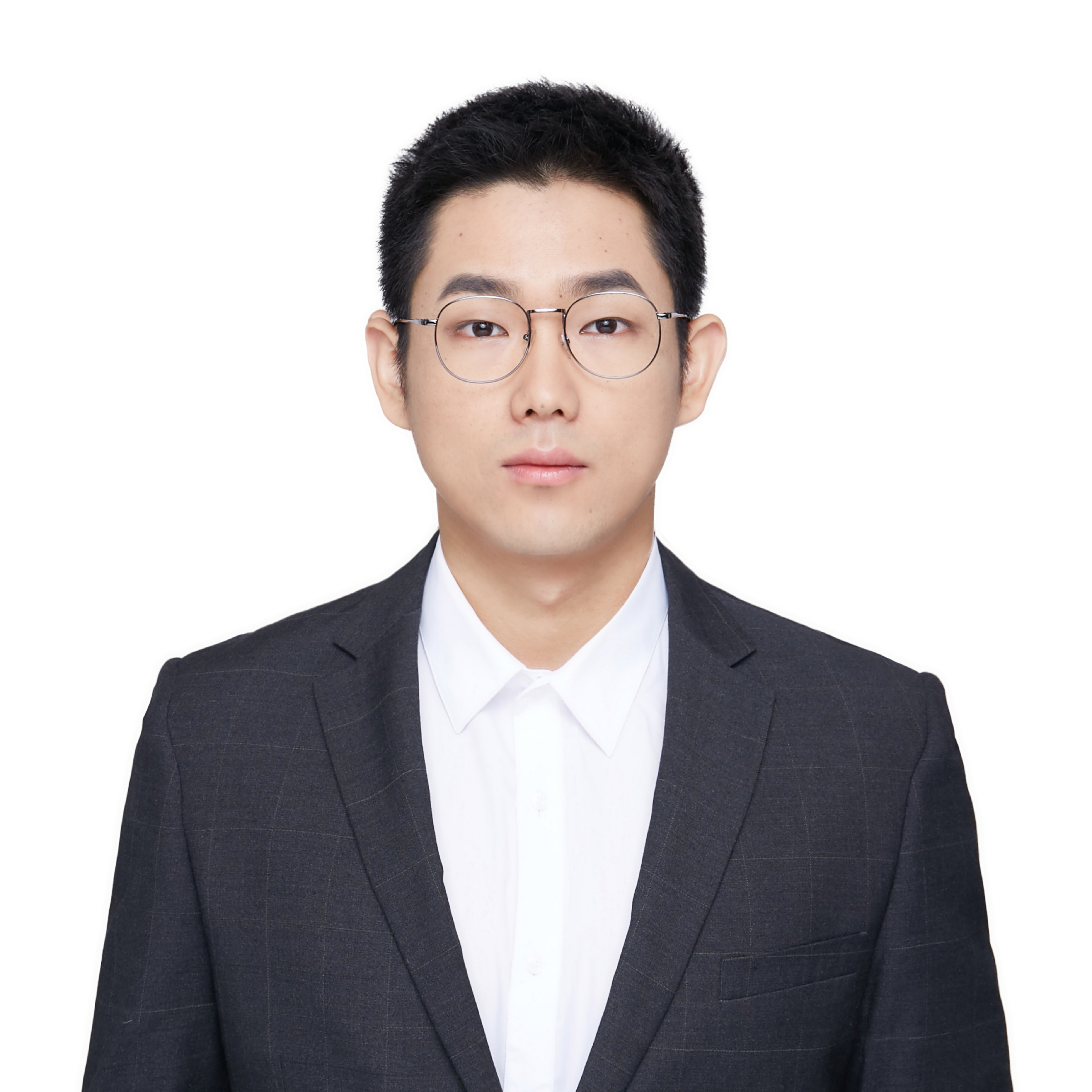}}]{Chenwei Wang}
received the B.S. degree from the School of Electronic Engineering, University of Electronic Science and Technology of China (UESTC), Chengdu, China, in 2018. He is currently pursuing the Ph.D. degree with the School of Information and Communication Engineering, University of Electronic Science and Technology of China, Chengdu, China.
His research interests include radar signal processing, machine learning, and automatic target recognition.
\end{IEEEbiography}

\begin{IEEEbiography}[{\includegraphics[width=1in,height=1.25in,clip,keepaspectratio]{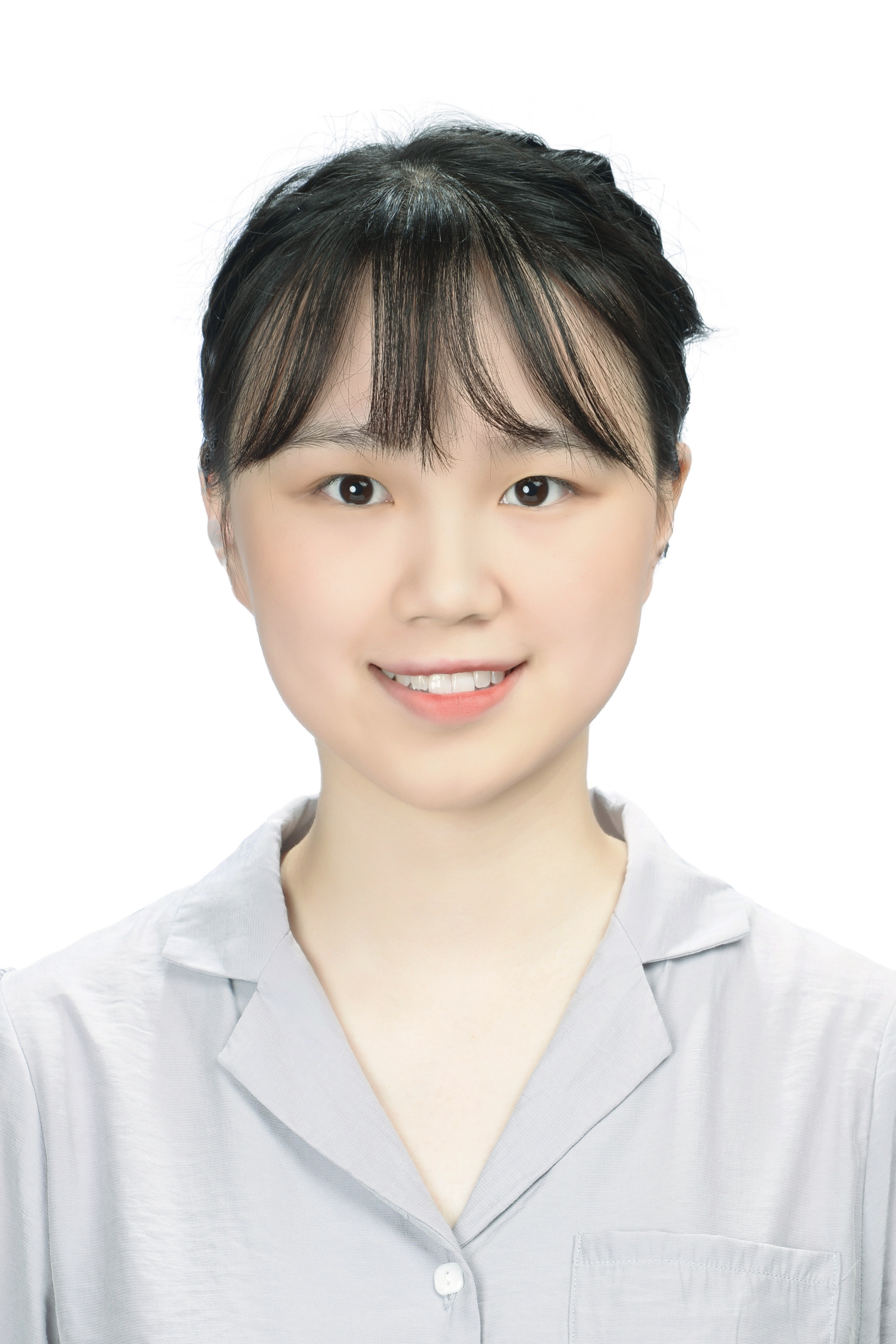}}]{Siyi Luo}
received the B.S. dual degree from Chongqing University (CQU), Chongqing, China, in 2021. She is currently pursuing the M.S. degree with the School of Information and Communication Engineering, University of Electronic Science and Technology of China, Chengdu, China. Her research interests include machine learning, target detection, and automatic target recognition.
\end{IEEEbiography}

\begin{IEEEbiography}[{\includegraphics[width=1in,height=1.25in,clip,keepaspectratio]{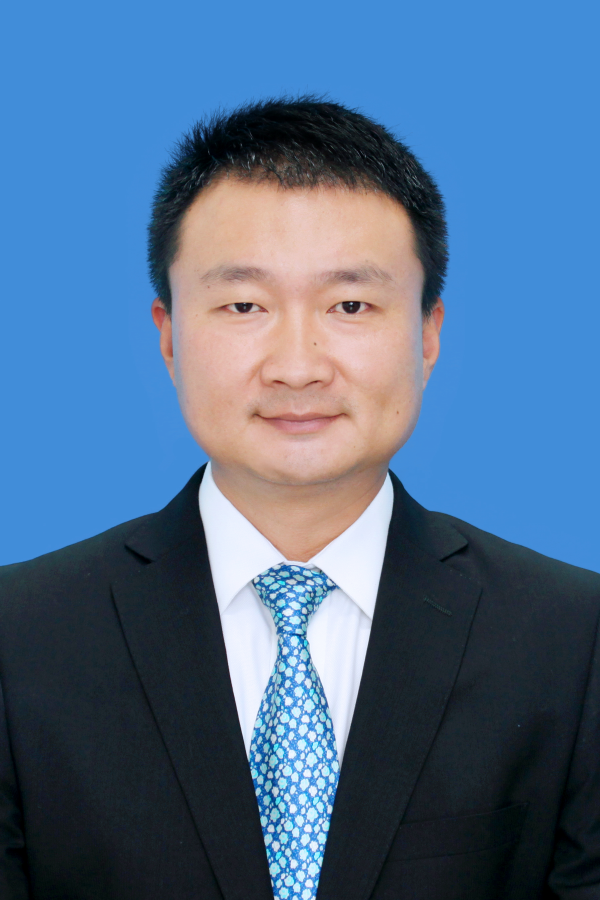}}]{Yulin Huang}
(M'08) received the B.S. and Ph.D. degrees from the School of Electronic Engineering, University of Electronic Science and Technology of China (UESTC), Chengdu, China, in 2002 and 2008, respectively. He is currently a Professor at the UESTC. His research interests include radar signal processing and SAR automatic target recognition.
\end{IEEEbiography}

\begin{IEEEbiography}[{\includegraphics[width=1in,height=1.25in,clip,keepaspectratio]{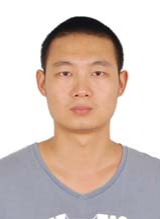}}]{Jifang Pei}
(M'19) received the B.S. degree from the College of Information Engineering, Xiangtan University, Hunan, China, in 2010, and the M.S. degree from the School of Electronic Engineering, University of Electronic Science and Technology of China (UESTC), Chengdu, China, in 2013. He received the Ph.D. degree from the School of Information and Communication Engineering, UESTC, in 2018. From 2016 to 2017, he was a joint Ph.D. Student with the Department of Electrical and Computer Engineering, National University of Singapore, Singapore. He is currently an Associate Research Fellow with the School of Information and Communication Engineering, UESTC. His research interests include radar signal processing, machine learning, and automatic target recognition.
\end{IEEEbiography}

\begin{IEEEbiography}[{\includegraphics[width=1in,height=1.25in,clip,keepaspectratio]{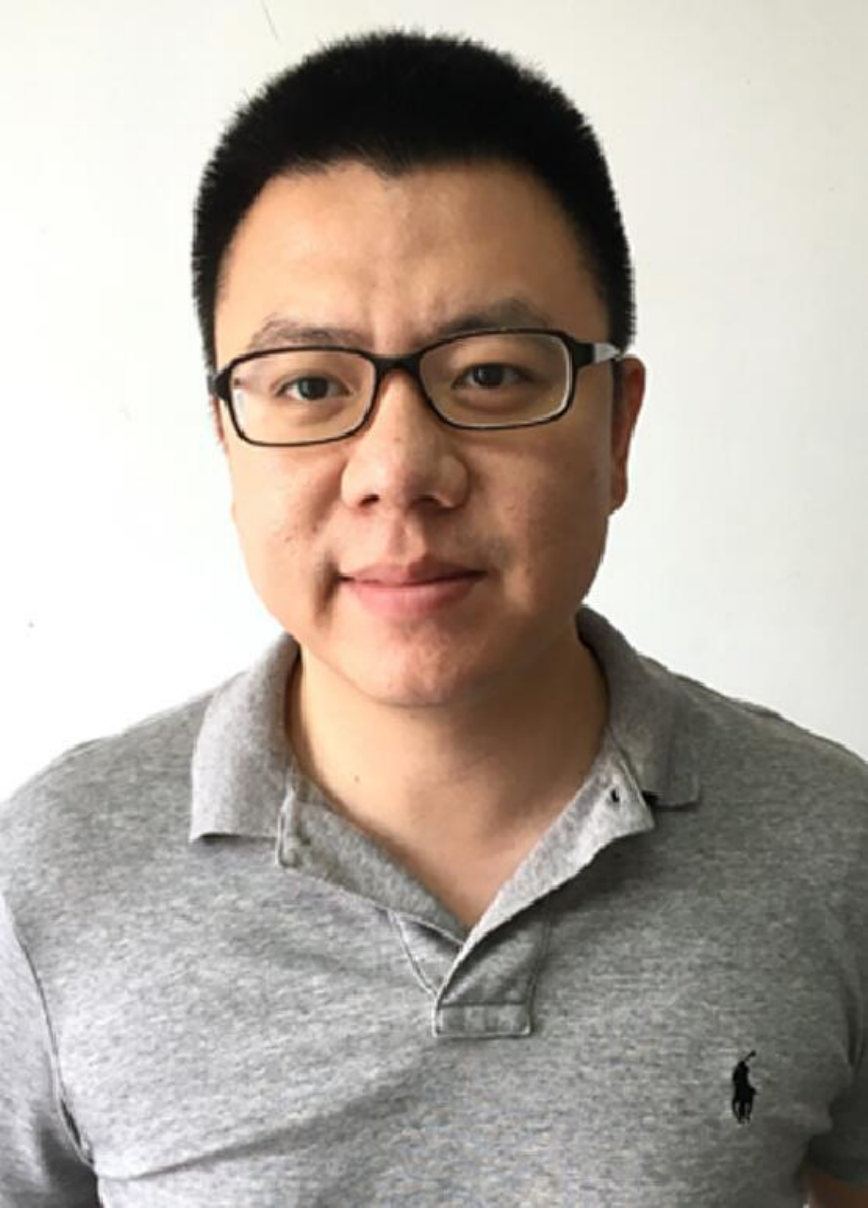}}]{Yin Zhang}
(M'16) received the B.S. and Ph.D. degrees from the School of Electronic Engineering, University of Electronic Science and Technology of China (UESTC), Chengdu, China, in 2008 and 2016, respectively. From September 2014 to September 2015, he had been a Visiting Student with the Department of Electrical and Computer Engineering, University of Delaware, Newark, USA. He is currently an Associate Research Fellow at the UESTC. His research interests include signal processing and radar imaging.
\end{IEEEbiography}

\begin{IEEEbiography}[{\includegraphics[width=1in,height=1.25in,clip,keepaspectratio]{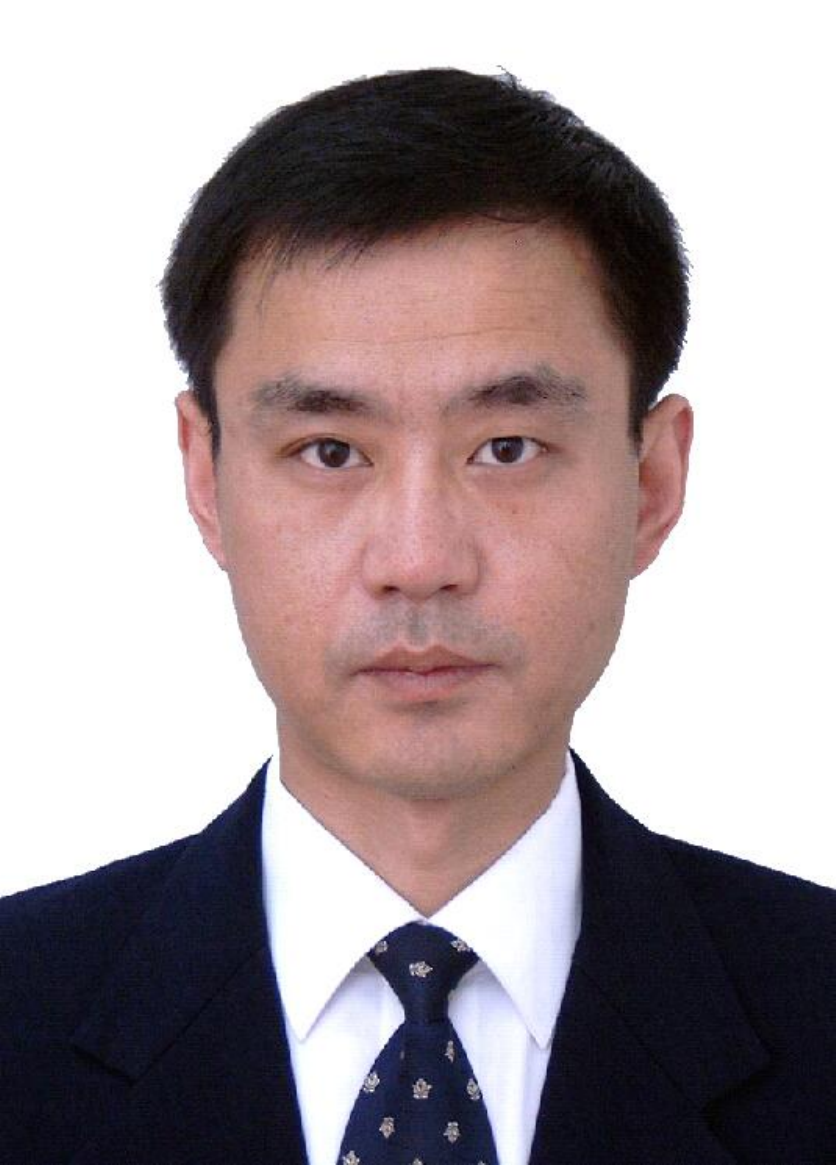}}]{Jianyu Yang}
(M'06) received the B.S. degree from the National University of Defense Technology, Changsha, China, in 1984, and the M.S. and Ph.D. degrees from the University of Electronic Science and Technology of China (UESTC), Chengdu, China, in 1987 and 1991, respectively. He is currently a Professor at the UESTC. His research interests include synthetic aperture radar and statistical signal processing. Prof. Yang serves as a Senior Editor for the Chinese Journal of Radio Science and the Journal of Systems Engineering and Electronics.
\end{IEEEbiography}


%






\end{document}